\documentclass[10pt,twocolumn,letterpaper]{article}

\usepackage{iccv}
\usepackage{times}
\usepackage{epsfig}
\usepackage{graphicx}
\usepackage{amsmath}
\usepackage{amssymb}
\usepackage{cite}
\usepackage{booktabs}
\usepackage{xcolor,colortbl}
\usepackage{multirow}
\usepackage{makecell}
\usepackage{adjustbox}
\usepackage{subfigure}

\definecolor{shadecolor}{rgb}{0.92,0.92,0.92}

\newcommand{\cmark}{\text{\ding{51}}}%
\newcommand{\xmark}{\text{\ding{55}}}%

\makeatletter %
\@namedef{ver@everyshi.sty}{}
\makeatother
\usepackage{tikz}

\newcommand{\name}{0}
\newcommand{\h}{0}
\newcommand{\w}{0.15}
\newcommand{\wa}{0.15}
\newlength \g
\usepackage[symbol]{footmisc}
\usepackage{pifont}%

\usepackage[pagebackref=true,breaklinks=true,letterpaper=true,colorlinks,bookmarks=false]{hyperref}

\usepackage[capitalize]{cleveref} 

\iccvfinalcopy 

\def\iccvPaperID{2817} 

\ificcvfinal\fi

\begin{document}

\title{MSFA-Frequency-Aware Transformer for Hyperspectral Images Demosaicing}

\author{$\text{Haijin Zeng}^1$, $\text{Kai Feng}^2$, $\text{Shaoguang Huang}^3$, $\text{Jiezhang Cao}^4$, $\text{Yongyong Chen}^5$, \\ $\text{Hongyan Zhang}^6$, $\text{Hiep Luong}^1$, and 
	$\text{Wilfried Philips}^1$\\
$^1\text{IMEC-IPI-UGent}$, $^2\text{NWPU}$, $^3\text{CUG}$, $^4\text{ETH Zurich}$, $^5\text{HIT}$, $^6\text{WHU}$ \\

{\tt\small haijin.zeng@ugent.be}
}

\maketitle
\ificcvfinal\thispagestyle{empty}\fi

\begin{abstract}
Hyperspectral imaging systems that use multispectral filter arrays (MSFA) capture only one spectral component in each pixel. 
Hyperspectral demosaicing is used to recover the non-measured components. 
While deep learning methods have shown promise in this area, they still suffer from several challenges,
including limited modeling of non-local dependencies, lack of consideration of the periodic MSFA pattern that could be linked to periodic artifacts, and difficulty in recovering high-frequency details. To address these challenges, this paper proposes a novel demosaicing framework, the MSFA-frequency-aware Transformer network (FDM-Net). FDM-Net integrates a novel MSFA-frequency-aware multi-head self-attention mechanism (MaFormer) and a filter-based Fourier zero-padding method to reconstruct high pass components with greater difficulty and low pass components with relative ease, separately. The advantage of Maformer is that it can leverage the MSFA information and non-local dependencies present in the data. Additionally, we introduce a joint spatial and frequency loss to transfer MSFA information and enhance training on frequency components that are hard to recover. Our experimental results demonstrate that FDM-Net outperforms state-of-the-art methods with 6dB PSNR, and reconstructs high-fidelity details successfully.

\end{abstract}

\section{Introduction}
%
%
%
%
Hyperspectral imaging (HI) captures light across a broad range of spectral bands, including those within the visible and beyond near-infrared spectrum. This provides much higher spectral resolution than the 3 spectra, leading to more accurate material characterization than is achievable through RGB imaging. This capability makes HI a valuable tool in numerous fields, including medical imaging, astronomy, food quality control, remote sensing, precision agriculture and pharmaceuticals. \cite{2021ICCV_HSIDenoising,zhang2019hyperspectral,chang2007hyperspectral,cao2016computational}.

\begin{figure}[!t]
\centering
\includegraphics[width=3.2in]{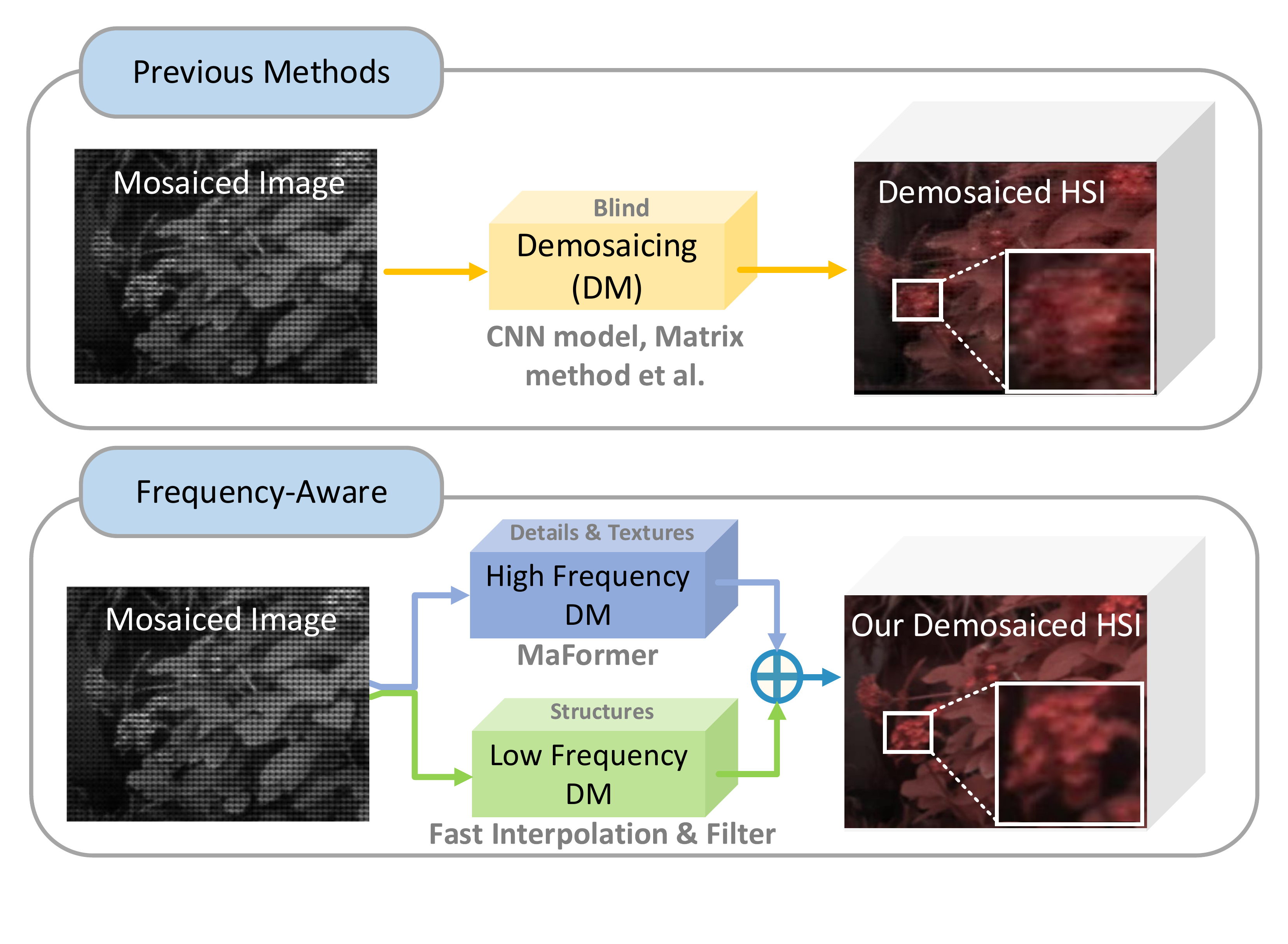}
\caption{Overview of previous methods and our frequency-aware HSI demosaicing framework. In contrast to current demosaicing methods that do not differentiate between the facile low-pass components and arduous high-pass details, we propose a frequency-aware demosaicing framework, which employs a customized transformer to reconstruct the hard high-pass components and data-independent but stable traditional interpolation-filtering to recover low-pass parts expeditiously. The proposed approach yields a significant improvement in the reconstruction of details.}
\label{fig_demo_1stpage}
\end{figure}


However, their employment in computer vision is limited due to slow acquisition times attributed to spatial or spectral scanning. To address this issue, snapshot HI systems \cite{Spectral_Imaging_review,arad2022ntire}, such as computed tomography \cite{dalton2021monte,koundinyan2018material} and light-field imaging \cite{light_field_2,light_field_1}, have been introduced recently, which capture both spectral and spatial information rapidly. These snapshot HI systems can be realized by snapshot mosaic HI systems or Multi-Spectral Filter Array (MSFA) cameras \cite{lapray2014multispectral}. The latter uses an MSFA to acquire spectral information in a single 2D image sensor exposure, similar to RGB cameras. However, MSFA cameras employ larger Color Filter Arrays (CFAs), such as 3 × 3, 4 × 4, or 5 × 5 \cite{detailed_HIS}.


The availability of MSFA cameras, designed with tiny Fabry-Pérot interferometry filters on top of CMOS or InGaAs sensors to obtain wavelength selectivity via a multiple-beam interference process, has been increasing for researchers and professionals at more accessible prices. 
Prominent examples of such cameras include the IMEC SNAPSHOT, XIMEA Snapshot USB3, and silios CMS series \cite{arad2022ntire}. However, to make optimal use of the spatial and spectral information provided by MSFA cameras, it is necessary to apply effective spectral demosaicing methods that can estimate a fully defined hyperspectral image (HSI).
Demosaicing large MSFAs presents a challenge due to the larger mosaic pattern and weaker inter-channel correlation in comparison to Bayer filter cameras. Although several demosaicing methods have been proposed, they exhibit limited demosaicing capability for high frequency details, resulting in the persistence of periodic artifacts. This may be due to the current CNN-based methods inadequately accounting for long-range dependencies \cite{cai2022degradation} and MSFA periodic information that is also critical for HSI demosaicing.

\emph{Motivation of Using Transformer:}
Specifically, during the MSFA imaging process, the entire spectral domain information is sampled and compressed into a single band, resulting in spatial-spectral confusion. Specifically, the nearest neighbors with similar spectral information are stored in a periodic MSFA pattern, causing them to be several pixels away in memory compared to RGB. This confusion occurs both across adjacent bands and throughout the entire spectral domain, and current CNN-based methods are unable to eliminate it. Non-local similarity has been identified as a critical factor in addressing spatial-spectral confusion  \cite{wang2022spatial,liu2018rank}. However, the receptive field of convolution limits its ability to leverage non-local information. In contrast, the Transformer architecture can exploit long-range non-local similarity and significantly improve reconstruction outcomes \cite{devlin2018bert,rao2021msa,transformer_pretrained,transformer_2,transformer_pyramid,UTransformer,transformer_swinir,liu2021swin}. Additionally, current methods struggle with detail recovery, while the Transformer has demonstrated exceptional capability in detecting subtle spatial differences \cite{dstransformer_HSI_res,he2022transfg,Restormer}.

Motivated by these observations and the inherently high-frequency nature of details in HSI, we propose an efficient HSI demosaicing network that employs a Transformer model and models MSFA information. Our method reconstructs the high-pass and low-pass components of HSI separately. Firstly, we utilize a Fourier zero-padding-based low-pass filter to quickly reconstruct the low-pass components that are easier to recover. Secondly, we introduce a novel MSFA-Frequency-aware Transformer, named \emph{MaFormer}, which focuses on the hard high-frequency details by concurrently modeling non-local dependencies and MSFA information. This enables us to recover the high-frequency details with reduced artifacts, as illustrated in Fig.~\ref{fig_demo_1stpage}. Finally, we integrate a joint spatial-frequency regularization term into the network, which utilizes both the MSFA pattern and frequency information to improve the reconstruction of details while preserving the fidelity of the low-pass components.
In summary, our contributions are three-fold:
\begin{enumerate}
    \item We propose a novel MSFA-frequency-aware HSI demosaicing framework that amalgamates the benefits of traditional methods with transformer to reconstruct HSI with precise details and fewer artifacts.
    \item By simultaneously incorporating non-local and MSFA periodic modeling, we present MaFormer, a tailored transformer designed specifically to demosaic the challenging high-pass HSI.
    \item Our FDM-Net outperforms state-of-the-art methods by a large margin, and produces highly accurate details.
\end{enumerate}

\section{Related Work}

\subsection{HSI Demosaicing}


 

Research on multispectral image demosaicing has been conducted in various studies \cite{WB,BTES,PPID,GRMR,Demosaic_matrix_completion,SpNet,DesNet,InNet,MCAN}. 
Current methods for demosaicing can be categorized into interpolation-based, matrix-factorization/recovery-based, and deep learning approaches. Interpolation and matrix-based methods \cite{WB,BTES,GRMR,PPID,Demosaic_matrix_completion} rely on spectral-spatial priors to reconstruct missing spectral and spatial information. This paper will focus on the latest learning-based approaches for HSI demosaicing.

CNNs has gained widespread popularity in various low-level image processing tasks, including image deblurring \cite{kupyn2019deblurgan, Restormer, cho2021rethinking}, denoising \cite{chen2021hinet, zeng_DPLRTA}, and super-resolution \cite{cao2022towards, wei2021unsupervised}. Although CNNs have been effectively employed in demosaicing \cite{ehret2019joint, liu2020joint, xing2021end}, their application has been predominantly limited to the Bayer pattern, which has a predominant green band. In contrast, spectral demosaicing necessitates the representation of multispectral correlations to enable CNN utilization. Consequently, researchers have introduced four distinct methods, namely DsNet \cite{DesNet}, SpNet \cite{SpNet}, In-Net \cite{InNet}, and MCAN \cite{MCAN}.
In particular, InNet \cite{InNet} applies a deep network employing a bilinear interpolated MSI cube as input, MCAN \cite{MCAN} proposes an end-to-end network that models joint spatial-spectral correlations in mosaic images. 
However, their capability to restore high frequency details is still restricted. Additionally, these methods do not account for long-range dependencies.


\subsection{Vision Transformer}
As an dominated architecture in NLP, the Transformer \cite{vaswani2017attention} is designed for sequence modeling, by incorporating a self-attention mechanism. 
It also has demonstrated remarkable performance in various vision-related tasks \cite{transformer_1,transformer_2,liu2021swin,transformer_swinir}. 
However, the use of transformer in image restoration \cite{transformer_pretrained,transformer_video} often involves dividing the input image into small, fixed-sized patches and processing each patch independently, which leads to two main issues \cite{liu2021swin,cao2023swin_unet}. Firstly, the restored image may exhibit border artifacts around the edges of each patch, and secondly, border pixels in each patch lose information that could have otherwise contributed to better restoration. 
Recently, the Swin Transformer \cite{liu2021swin} has emerged as a promising solution, incorporating the benefits of both CNNs and Transformers: on the one hand, it inherits the advantage of CNNs in processing large images due to its local attention mechanism; on the other hand, it retains the capability of Transformers in modeling long-range dependencies using the shifted window scheme \cite{transformer_swinir,cao2023swin_unet,liu2021swin}.


\begin{figure}[!t]
\centering
\includegraphics[width=3.2in]{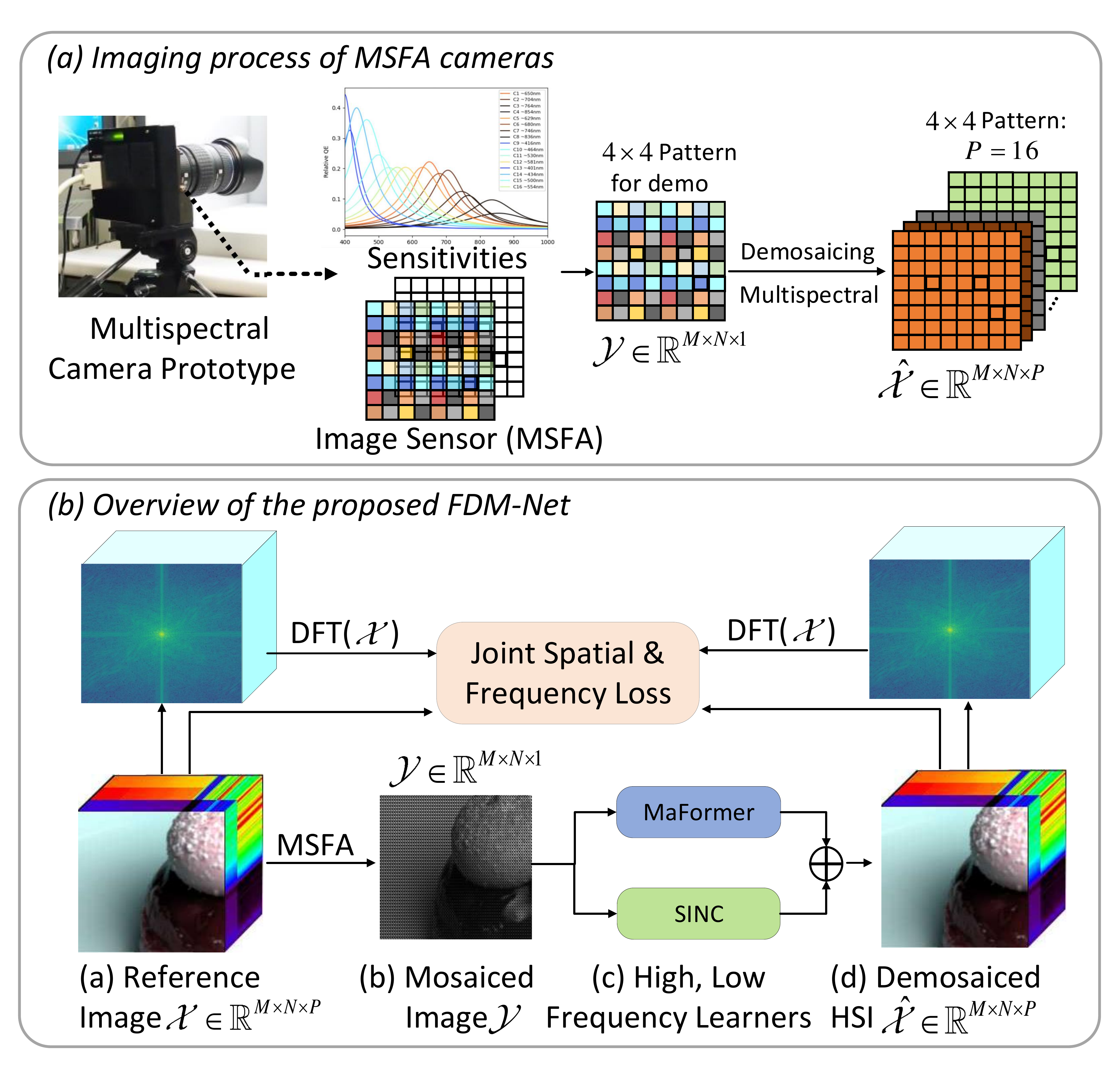}
\vspace{-2mm}
\caption{Overview of the imaging process of MSFA camera and our frequency-aware demosaicing framework: FDM-Net.}
\label{fig_sketch_1}
\end{figure}

\section{Method}

\subsection{Frequency-Aware Demosaicing Network} \label{model:overview}

The proposed MSFA-frequency-aware demosaicing network (FDM-Net) for HSI is depicted in Fig.~\ref{fig_sketch_1} (b).
The method was inspired by the recognition that while low pass structural information can be efficiently reconstructed by most demosaicing techniques, the main challenge lies in the recovery of high pass detail and texture information.
However, previous methods have not sufficiently differentiated between these two components and instead use a single integrated model to reconstruct both high and low pass components simultaneously.
To address this issue, we first decompose the HSI cube into its high pass and low pass components. Then, we customise a MSFA-Conv based Swin Transformer network (MaFormer) by performing non-local and MSFA periodic modeling simultaneously, and a sinc-interpolation block to reconstruct the high-low pass components.
Finally, we merge the reconstructed high and low pass components to obtain the final demosaiced HSI.

Specifically, given a mosaiced image $\mathcal{Y} \in \mathbb{R}^{M \times N}$, where $M$, $N$ are the image height and width, respectively.
Firstly, the low pass components $\hat{\mathcal{X}}_{LF}$ are reconstructed using Fourier
zero-padding (Lanczos windowed sinc \cite{duchon1979lanczos}) with guided-filter, which is a low-pass filter very accurate on smooth data, 
\begin{equation}
    \hat{\mathcal{X}}_{LF} = \operatorname{Filter}(\operatorname{Sinc}(\mathcal{Y}), \operatorname{Sinc}(\mathcal{Y})(:,:,0))\in \mathbb{R}^{M \times N \times C},
\end{equation}
where $u(x,y) = \sum_{m,n}v_{m,n}\emph{sinc}(x-m),\emph{sinc}(y-n)$ is the Sinc interpolation of $v(x,y)$, and $\emph{sinc}(t):=\omega_t\emph{sin}(\pi t)/(\pi t)$ for $t \neq 0$, $\emph{sinc}(0):=1$, $\omega_t=\frac{n}{\pi t}\emph{sin}(\pi t/n)$, if $\|t\|<n$.
Subsequently, we propose a customized transformer for reconstructing the high pass details $\hat{\mathcal{X}}_{HF}$. 

\begin{figure}[!t]
\scriptsize
\centering
\rotatebox{90}{\vspace{9mm}  \centering \small Image}
\includegraphics[width=0.74in]{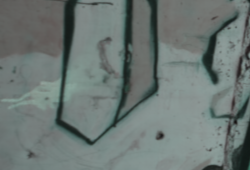}
\hspace{-0.5mm}
\includegraphics[width=0.735in]{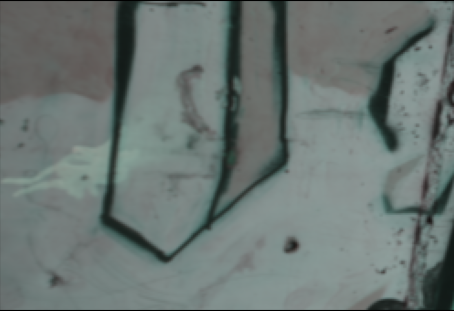}
\hspace{-0.5mm}
\includegraphics[width=0.74in]{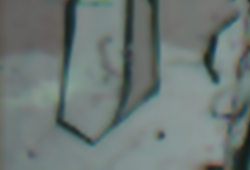}
\hspace{-0.5mm}
\includegraphics[width=0.74in]{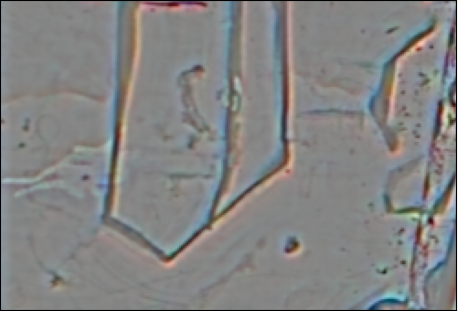}\\
\vspace{-1.5mm}
\rotatebox{90}{\vspace{9mm}  \centering \small Spectrum}
\subfigure[FDM-Net]{\includegraphics[width=0.76in]{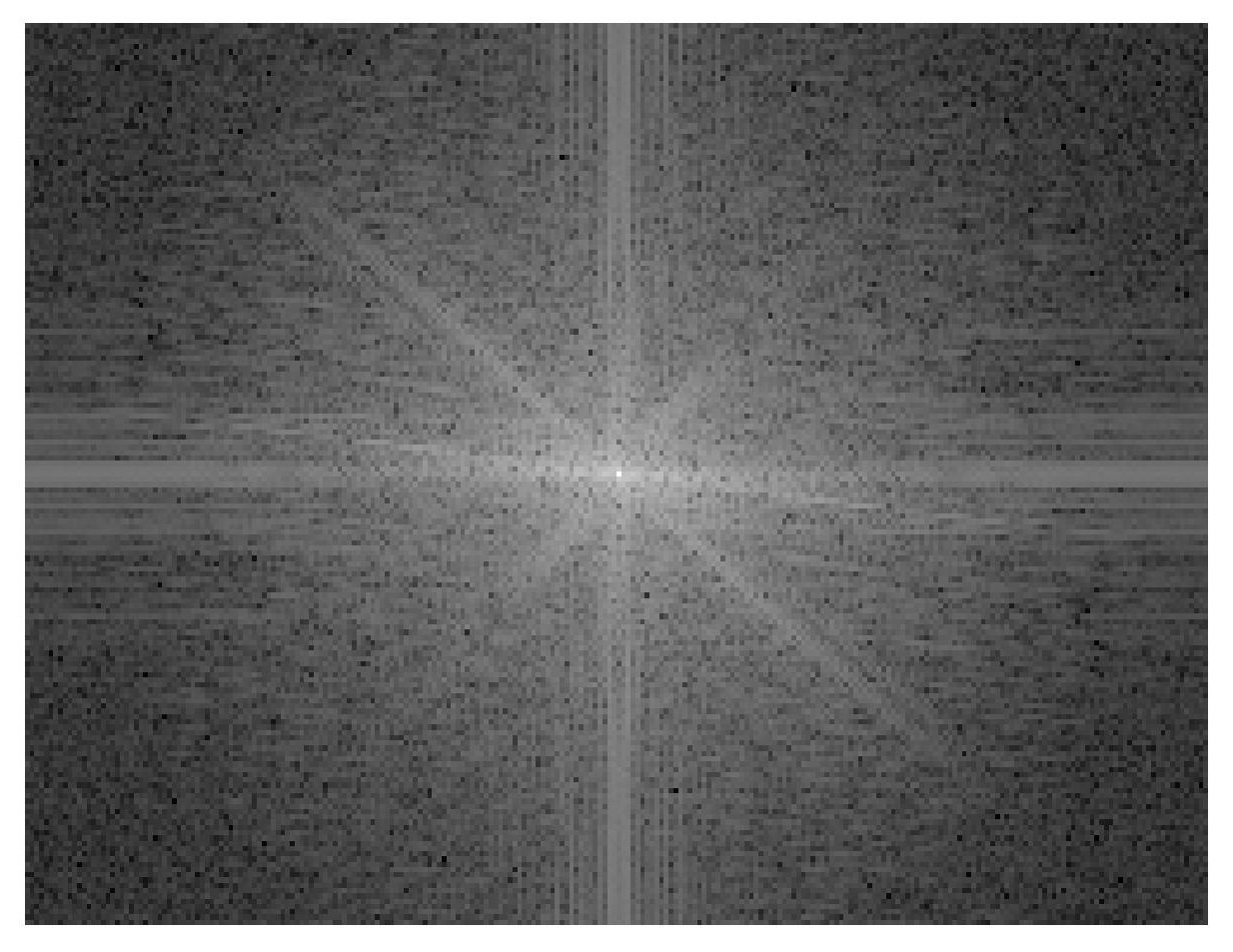}}
\hspace{-1mm}
\subfigure[GT]{\includegraphics[width=0.76in]{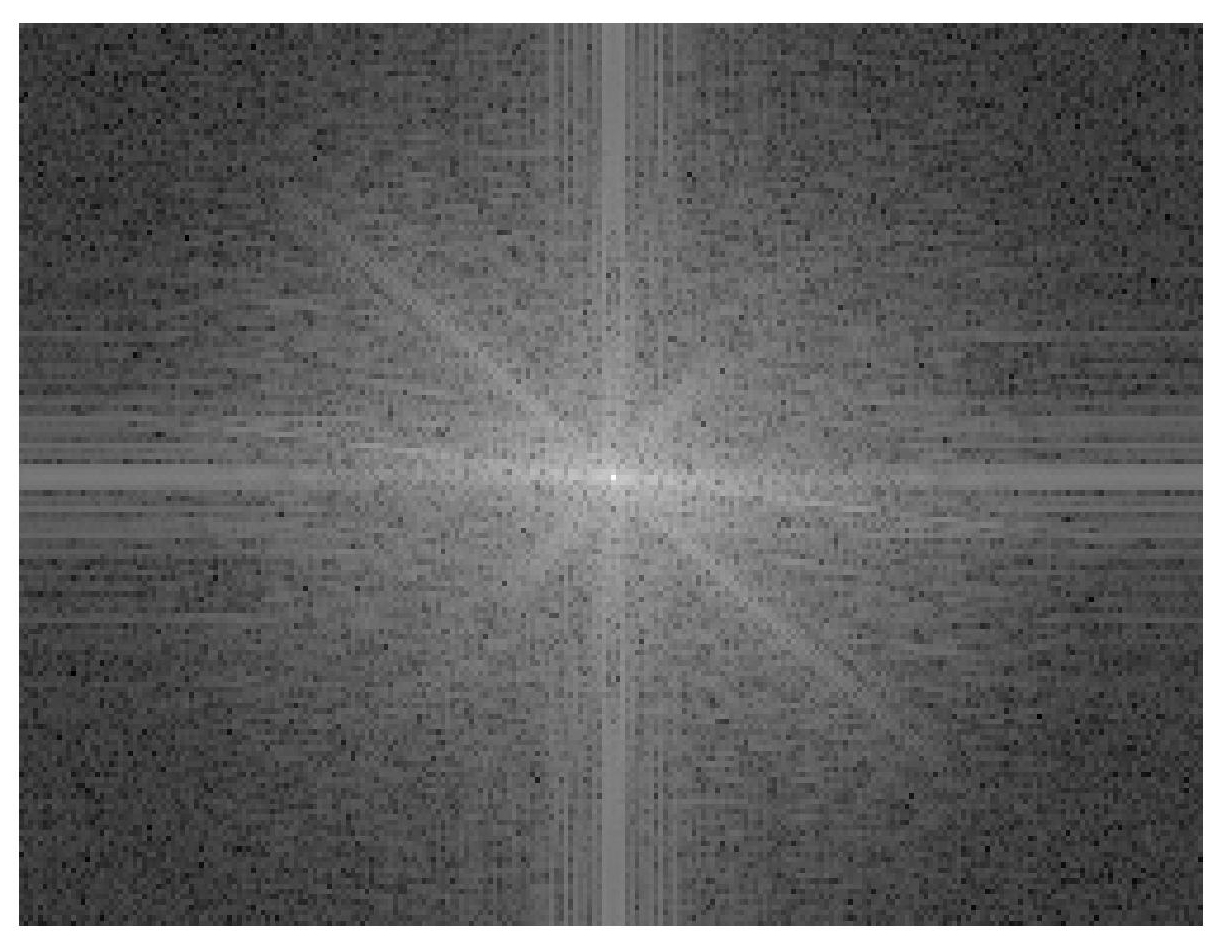}}
\hspace{-1mm}
\subfigure[Low-Pass]{\includegraphics[width=0.765in]{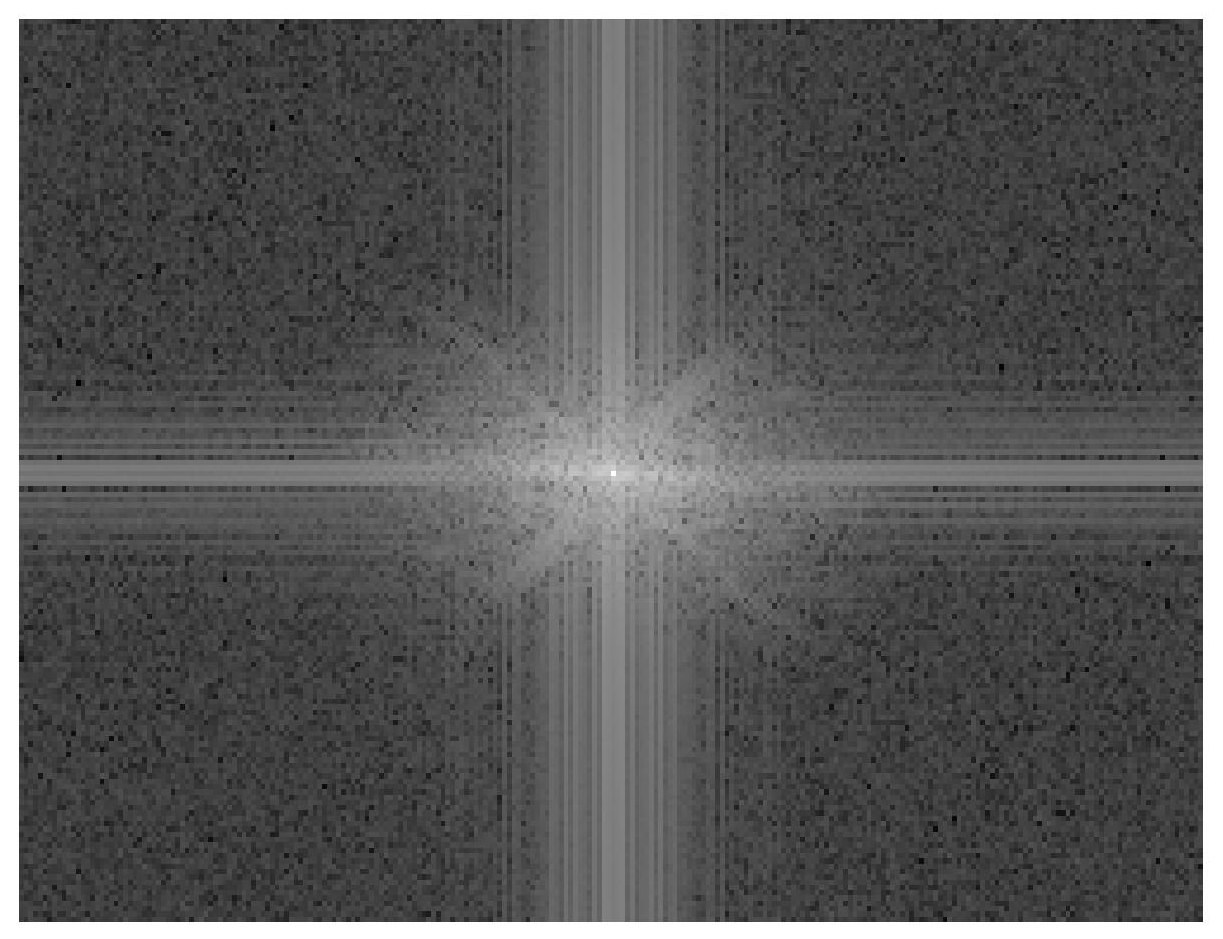}}
\hspace{-1mm}
\subfigure[High-Pass]{\includegraphics[width=0.765in]{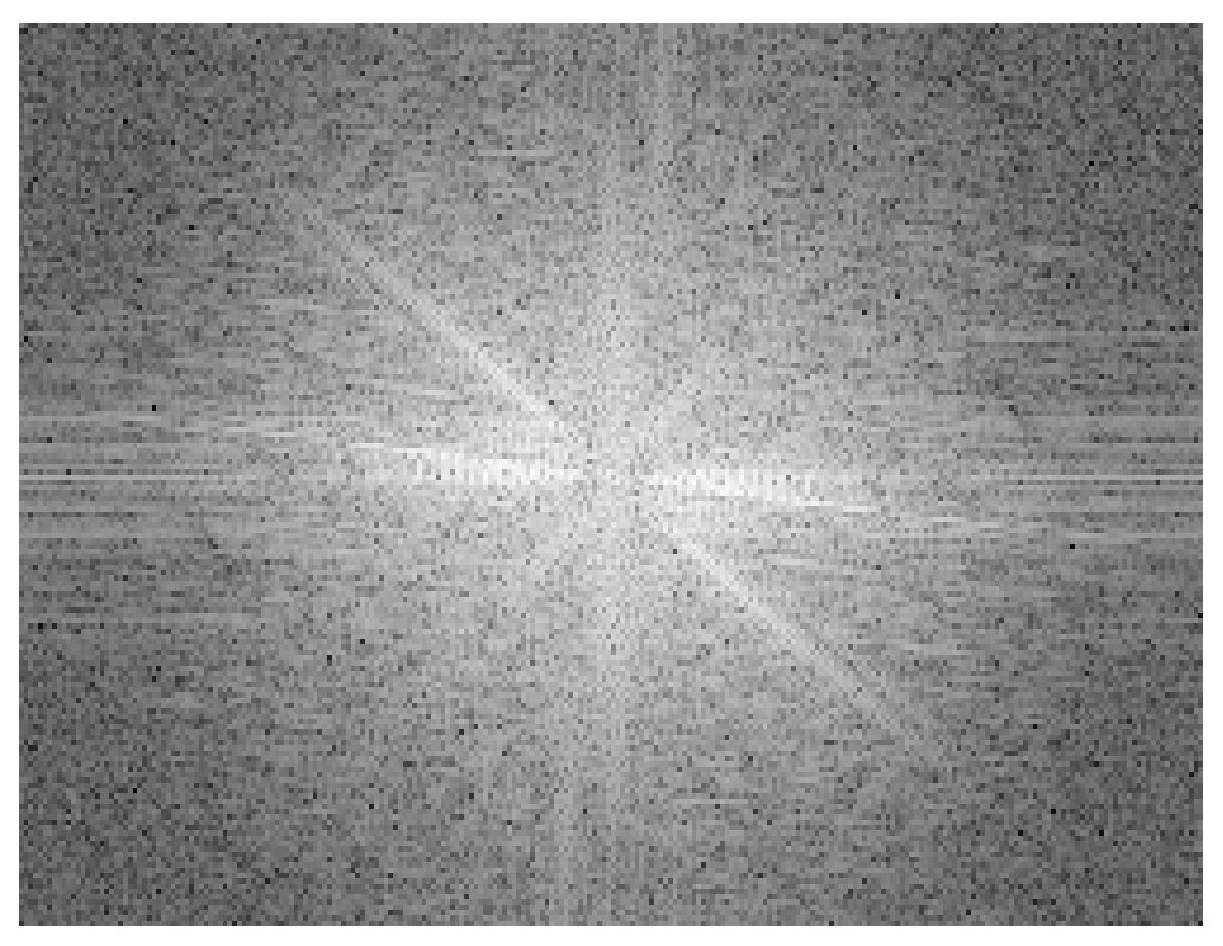}}
\vspace{-1.5mm}
\caption{Illustration of the low pass part and high pass component reconstructed by our FDM-Net. The output of our FDM-Net is generated by adding the low pass and high pass it reconstructed.}
\label{fig_HL_demo}
\end{figure}

Our primary objective is to develop an effective and efficient module for the recovery of high pass details and textures, which presents a significant challenge. To address this, we select the transformer network, as it has demonstrated outstanding performance in distinguishing even subtle spatial differences by characterizing sequential spatial data \cite{he2022transfg,da2022transformer}. Fig.~\ref{fig_HL_demo} shows that the reconstructed low pass component contains clear smoothed structural information, while the high pass component learned by our MaFormer effectively captures the details and textures of the HSI. The demosaiced HSI, which is a result of the aggregation of the high pass and low pass components produced by our FDM-Net, is shown to be highly similar to the ground truth.

\begin{figure*}[!t]
\centering
\includegraphics[width=6.5in]{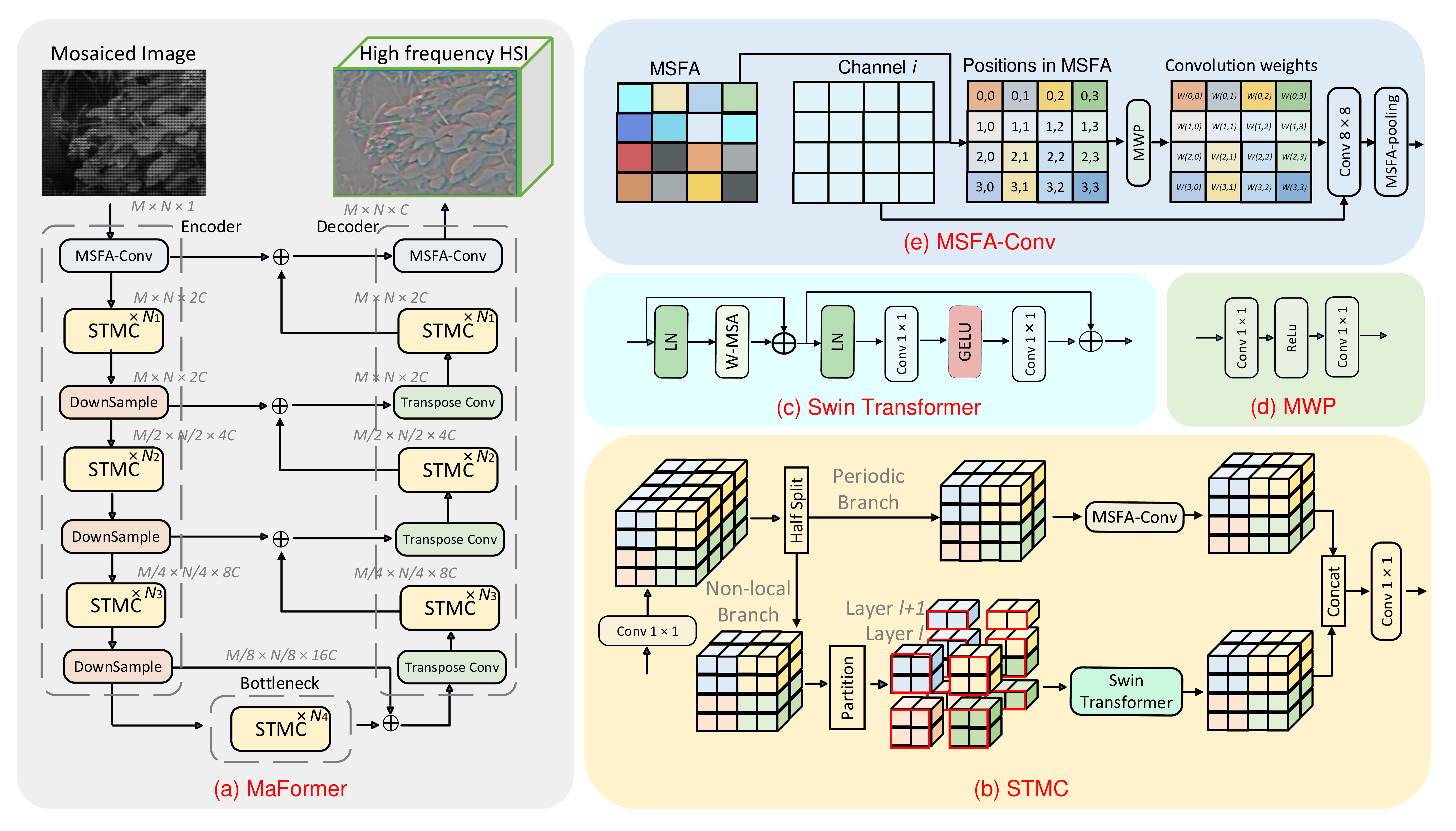}
\vspace{-3mm}
\caption{The architecture of our MaFormer. (a) MaFormer consists of an encoder, a bottleneck, and a decoder. MaFormer is built up by STMCs. (b) STMC
is composed of a parallel group convolution, which includes a periodic branch and a non-local branch. (c) The swin transformer used in non-local branch. (d) The weights prediction block of MSFA-Conv. (e) The MSFA-driven convolution: MSFA-Conv.}
\label{fig_sketch_2}
\end{figure*}

\subsection{Customized High Frequency Transformer} \label{maformer}
The proposed \emph{MaFormer} is the cornerstone of our FDM-Net, which adopts an overall architecture resembling a U-Net, as illustrated in Fig. \ref{fig_sketch_2} (a). Comprising an encoder, a bottleneck, and a decoder, \emph{MaFormer} employs downsampling and upsampling techniques through transpose convolutions. This architectural choice, which differs from stacking modules layer by layer without scaling, has been shown to enhance the performance of the algorithm and increase the receptive field of the basic CNN and the proposed MSFA convolution, as detailed in Sec. \ref{MSFA_SW}. However, downsampling inevitably leads to a loss of information, which we address by incorporating residual connections between the encoder and decoder stages.


Specifically, the first module of MaFormer is the MSFA-Conv, as illustrated in Fig. \ref{fig_sketch_2} (a). Its input is the raw mosaic data $\mathcal{Y} \in \mathbb{R}^{M \times N}$ that is sampled from the latent HSI $\mathcal{X} \in \mathbb{R}^{M \times N \times C}$. Here, $M$ and $N$ denote the height and width of the observed raw data, respectively, and $C$ denotes the number of channels.
Firstly, the MSFA-Conv extracts feature maps $\mathcal{X}_0 \in \mathbb{R}^{M \times N \times 2C}$ from $\mathcal{Y}$.
Secondly, the $\mathcal{X}_0$ is fed into three paired STMC and Downsample blocks, resulting in hierarchical feature maps. The Downsample layer is implemented using a $2 \times 2$ convolution without bias, which generates a downsampled feature map with double channels while half spatial resolution. We denote the outputs of these three paired STMC and Downsample groups as $\mathcal{X}_i, i=1, 2, 3$, respectively.
Thirdly, the bottleneck processes $\mathcal{X}_3$ using a pure STMC without any sampling.
Subsequently, a symmetric decoder is designed as a classical U-Net. It also consists of three STMC blocks and an MSFA Conv, but the downsample layers are replaced with transpose convolutions, which are used to upsample the spatial dimensions of intermediate feature maps.
Finally, the high-frequency information, such as details and textures of the latent hyperspectral image, i.e., $\hat{\mathcal{X}}_{HF} \in \mathbb{R}^{M \times N \times C}$, is learned and reconstructed by an MSFA-Conv block.


\subsection{MSFA-based Half Swin Transformer} \label{MSFA_SW}
The crucial component of the proposed \emph{MaFormer} is the proposed integrated Swin-Tansformer \& MSFA-Conv (STMC) module. Fig.~\ref{fig_sketch_2} (b) illustrates the STMC module utilized to process the input tensor $\mathcal{X}_0 \in \mathbb{R} ^{M \times N \times 2C}$. Firstly, $\mathcal{X}_0$ is linearly projected via a $1 \times 1$ convolution layer, after which it is split into two sub-feature maps along the channel orientation,
\begin{equation}
    \mathcal{X}_0 = [ \mathcal{X}_0^{p}\in \mathbb{R}^{M \times N \times C}, \mathcal{X}_0^{nl} \in \mathbb{R}^{M \times N \times C} ].
\end{equation}
Then, $\mathcal{X}_0^{p}$ passes through the \emph{Periodic Branch} to model periodic MSFA information, while $\mathcal{X}_0^{nl}$ passes through the \emph{Non-local Branch} to model non-local dependencies.

\subsubsection{Periodic Branch}
The \emph{Periodic Branch} employs the \emph{MSFA-Conv} block to apply a MSFA-driven convolution operator, as shown in Fig. \ref{fig_sketch_2} (e). This operator refines the features based on the relative positions of the elements in input, which are determined by MSFA pattern. Specifically, for an element with index $(i, j)$, $p \times p$ MSFA pattern, its relative position is denoted as $\left(m, n\right)=(i \bmod p, j \bmod p)$. The relative position matrix $R$, with element $(m, n)$, is then fed into a MSFA attention weights prediction block (MWP), which consists of two $1 \times 1$ convolution layers and one $\operatorname{ReLU}$ activation function, as shown in Fig. \ref{fig_sketch_2} (d). The MWP generates a MSFA-driven convolution kernel with weight $W$ as follows:
\begin{equation}
W= \operatorname{Conv}1\times1(\operatorname{ReLU}(\operatorname{Conv}1\times1(R))).
\end{equation}
The kernel $W$ assigns the same weights to elements with the same relative positions, allowing neighboring elements sampled with the same wavelength to share similar spectral distributions. The resulting feature map is
\begin{equation}
    F_0^{p} = \operatorname{Conv}8\times8(\mathcal{X}, W) \in \mathbb{R}^{M \times N \times 2C},
\end{equation}
is obtained by convolving the input $\mathcal{X}$ with the kernel $W$ using an $8\times8$ convolution operation. This MSFA-Conv block is designed to effectively and efficiently model periodic information in the input.
Then, feature $F_0^{p}$ is aggregated by MSFA pooling, instead of using normal pooling, e.g., maximum or average pooling. 
Specifically, MSFA pooling aggregates the feature points of $F_0^{mc}$ with the same relative position to get $F_1^{p} \in \mathbb{R}^{\frac{M}{2} \times \frac{N}{2} \times 2C}$, where 
\begin{equation}
    F_1^{p}\left(i, j, k\right)=c \sum_{s=0}^{\frac{m}{4}-1} \sum_{t=0}^{\frac{n}{4}-1} F_0^{p}\left(i+4s, j+4t, k \right),
\end{equation}
$c=\frac{1}{m / 4 \times n / 4}$. After that, two $3 \times 3$ Residual Convolutional (RConv) are used to refine the feature map $F_1^{p}$,
\begin{equation}
   F_2^{p} = \operatorname{RConv}(\operatorname{RConv}(F_1^{p})).
   \label{eq:periodic_branch}
\end{equation}

\subsubsection{Non-local Branch}
The \emph{non-local branch} computes MSA \cite{vaswani2017attention} within position-specific shifted windows  and cross-window connection, by using swin transformer \cite{liu2021swin}.
Given an input $\mathcal{X}_0^{nl} \in \mathbb{R}^{M \times N \times C}$ split from $\mathcal{X}_0$, the \emph{non-local} branch partitions it into $K \times K$ local windows without overlapping, it reshapes it into $\mathcal{X}_0^{nl} \in \mathbb{R}^{\frac{MN}{K^2} \times K^2  \times C}$, where $K$ denotes the window size. 
To take into account the cross window connection, regular and shifted window partitioning are used alternately here \cite{liu2021swin}, as shown in Fig. \ref{fig_sketch_2} (b).
Then, the self-attention of each local windows $X_{nl} \in \mathbb{R}^{K^2 \times C}$ is computed, i.e., 
\begin{equation}
    Q_{nl} = X_{nl}W_Q, K_{nl} = X_{nl}W_K, V_{nl} = X_{nl}W_V, 
\end{equation}
where $Q\in \mathbb{R}^{K^2 \times d}, K\in \mathbb{R}^{K^2 \times d}, V\in \mathbb{R}^{K^2 \times d}$ are the \emph{query, key} and \emph{value}. 
$W_Q, W_K, W_V$ are projection matrices, which are shared across different partitioned local windows.
Then, the local self-attention matrix of local windows is computed as follows:
\begin{equation}
\text{Attention}(Q_{nl}, K_{nl}, V_{nl}) = \operatorname{SoftMax}(\frac{Q_{nl}K_{nl}^T}{\sqrt{d}}+B)V_{nl},  
\label{eq:nonlocal_branch}
\end{equation}
where $d$ is the \emph{query / key} dimension, $B \in \mathbb{R}^{K^2 \times K^2}$ is the learnable relative parameters depicts positional encoding.
Subsequently, the attention feature maps are fed into a LayerNorm(LN) layer, and then pass through two fully connected layers with GELU, i.e., MLP. 
In addition, the residual connection is also added to each input of LN, as shown in Fig. \ref{fig_sketch_2} (c), and the output is reshaped back to size $M \times N \times C$.
Finally, the outputs of \emph{periodic branch} in Eq. (\ref{eq:periodic_branch}) and \emph{non-local} branch Eq. (\ref{eq:nonlocal_branch}) are concatenated and then a fully connected layer is used to fuse the information between \emph{periodic branch}  and \emph{non-local branch}, generate the final output $X_1 \in \mathbb{R}^{M \times N \times 2C}$ of STMC. 
Following STMC, the $\operatorname{Downsample}$ operator samples $X_1$ with half spatial resolution and double channels.

Overall, our MaFormer combines the non-local modeling ability of Swin Transformer block and 
MSFA periodic modeling ability of MSFA-Conv. 
Furthermore, we enhance the integrated periodic and non-local modeling ability by stacking our STMC in a down-sample \& up-sample U-Net style, together with transpose convolution. 
In addition to take into account extra MSFA information, this is also computationally cheaper than global standard MSA, due to the split and concatenation operations within STMC can act as the group convolution with two groups.

\subsection{Joint Spatial and Frequency Loss} \label{joint_loss}

The demosaicing of HSI is an ill-posed inverse problem, where a single observed mosaic image can correspond to multiple HSIs. To mitigate this difficulty, as regularization-based optimization methods \cite{l12SSTV, zeng_DPLRTA}, we propose the incorporation of a joint spatial and frequency loss as a constraint in the optimization procedure, to decrease the potential solution space. 
\emph{Firstly}, consider that in low frequency reconstruction phase, our method leverages the use of classical interpolation to reconstruct the low frequency components quickly, and a guided filter to remove noise and refine the reconstructed low frequency parts.
To preserve the accuracy of the known high-frequency information, we introduce an MSFA-based $L_1$ loss to regularize the sampled pixels,
\begin{equation}
    L_1^s = \| x - \hat{x}\odot M\|_1.
\end{equation}
where $M$ is the MSFA sample mask, $x, \hat{x}$ are the ground truth and demosaiced HSI, respectively. 
$\odot$ denotes element-wise multiplication.

\emph{Secondly,} based on our observations, the lower frequency component within the high frequency part of the HSI is relatively easier to reconstruct, while the main challenge lies in the reconstruction of the higher frequency component, which contains complex details and textures.
To enhance the network's ability to model these challenging cases, we introduce the Focal Frequency Loss (FFL) \cite{focal_loss} instead of using frequency loss directly. The FFL focuses the network on the most challenging frequencies during its training,
\begin{equation}
    L_\mathrm{FFL}=\frac{1}{M N} \sum_{u=0}^{M-1} \sum_{v=0}^{N-1} w(u, v)\left|F_{\hat{x}}(u, v)-F_x(u, v)\right|^2 .
\end{equation}
where $M \times N$ is the image size, $(u, v)$ denotes the coordinate of a spatial
frequency on the frequency spectrum, $w(u, v)$ is the matrix element, i.e.,
the weight for the spatial frequency at $(u, v)$, it is defined as:
$$
w(u, v)=\left|F_{\hat{x}}(u, v)-F_x(u, v)\right|^\alpha,
$$
where $\alpha$ is the scaling factor for flexibility ($\alpha=1$), 
\begin{equation}
    F(u, v) = \frac{1}{M N} \sum_{u=0}^{M-1} \sum_{v=0}^{N-1} f(x, y) \cdot e^{-i2\pi(\frac{ux}{M}+\frac{vy}{N})}.
\end{equation}
The gradient through the spectrum weight matrix only serves as a weight for each frequency.
The Focal Frequency Loss (FFL) can be understood as a weighted average of the frequency differences between the reference and demosaiced images.
By using the FFL, the loss function is re-weighted to give priority to the reconstruction of the most complex details of textures with challenging frequencies, and to down-weight easier cases. In addition, the focus region is dynamically updated, which improves the immediate challenging frequencies and results in a gradual refinement of the generated images.

\emph{Subsequently,} $L_1$ loss is also added to the high frequency part, to ensure complete and global texture information in image domain can be learned by our network, i.e., 
\begin{equation}
    L_1^c = \| x - \hat{x}_l- \hat{x}_h\|_1
\end{equation}
where $\hat{x}_l$ and $\hat{x}_h$ are the predicted high frequency cube and low frequency part. 
\emph{Finally,} our joint frequency and spatial loss function for hyperspectral image demosaicing is formulated as follows:
\begin{equation}
    L = \alpha_1 L_1^s + \alpha_2 L_\mathrm{FFL} + \alpha_3 L_1^c 
\end{equation}
where $\alpha_1 = 0.1, \alpha_2 = 1,$ and $\alpha_3 = 1$.

\definecolor{Gray}{gray}{0.90}
\begin{table*}[!htbp]
\centering
\setlength{\tabcolsep}{1.2pt}
\renewcommand{\arraystretch}{1.2}%
\caption{Demosaicing results and running time compared with other methods. FDM-Net achieves SOTA results.}
\label{Table:PQI_1}
\scalebox{0.9}{
\begin{tabular}{cccccccccc}
\toprule[0.05em]
\rowcolor{Gray}
Datasets & Method &	WB \cite{WB} 	&	BTES \cite{BTES}	&	PPID \cite{PPID} & GRMR \cite{GRMR}  & In-Net \cite{InNet}& MCAN \cite{MCAN} & \textbf{FDM-Net} & \textbf{FDM-Net-L}\\
\midrule[0.05em]
\multirow{4}{*}{ARAD 901}  & PSNR $\uparrow$ &  29.27 & 29.52& 36.87 & 29.35 & 44.98& 41.60& \textbf{52.41} & 51.65\\
& SSIM $\uparrow$ & 0.956 & 0.947& 0.969 &0.960 &0.993 &0.988& \textbf{0.998} & 0.997\\
& SAM $\downarrow$  & 0.093&   0.089&   0.090&   0.096   &  0.011   &  0.020   &\textbf{ 0.004} & 0.004\\
& MRAE $\downarrow$ &-&-&-&-& 0.014 &0.022 & \textbf{0.005} & 0.005\\
\Xcline{1-10}{0.1pt}
\multirow{4}{*}{ARAD 903}  & PSNR $\uparrow$ &  30.95 & 31.06& 39.10 & 30.99& 41.09&36.92 & \textbf{48.78} & 48.63\\
& SSIM $\uparrow$ & 0.965 & 0.955& 0.977 &0.967 &0.985 &0.936 & \textbf{0.993} & 0.993\\
& SAM $\downarrow$ & 0.089 & 0.078 & 0.063 & 0.085 & 0.041 & 0.065& \textbf{ 0.011} & 0.011\\
& MRAE $\downarrow$ &-&-&-&-& 0.037& 0.071 & \textbf{0.012} & 0.012\\
\Xcline{1-10}{0.1pt}
\multirow{4}{*}{ARAD 907}  & PSNR $\uparrow$ &  33.85 & 33.49& 38.81 & 33.96& 43.50& 45.29 & \textbf{51.81} & 50.84\\
& SSIM $\uparrow$ & 0.943 & 0.929& 0.959 &0.944 &0.984 &0.991 & \textbf{0.997} & 0.997\\
& SAM $\downarrow$ & 0.113 & 0.134 & 0.079 & 0.113 & 0.033& 0.030& \textbf{0.012} & 0.013\\
& MRAE $\downarrow$ &-&-&-&-& 0.043 & 0.038 & \textbf{ 0.015} & 0.017\\
\Xcline{1-10}{0.1pt}
\multirow{5}{*}{50 HSIs} 
& PSNR $\uparrow$ & 31.17 & 30.94& 35.98 &31.38 &42.88  &43.22 & \textbf{49.23} & 48.60\\
& SSIM $\uparrow$ & 0.912 & 0.892 & 0.937& 0.922 & 0.981 &  0.986& \textbf{0.996} & 0.995\\
& SAM $\downarrow$ &0.158 & 0.176 & 0.121 & 0.150 & 0.034  &  0.034& \textbf{0.013} & 0.014\\
& MRAE  $\downarrow$ & - & - & - & - & 0.043  &0.044 & \textbf{0.017} & 0.018\\
\Xcline{1-10}{0.1pt}
Average & TIME (s)  & 0.11 & 0.12 & 0.75 & 15.18 &  0.014 & 0.015 & 0.054 & 0.029 \\
\bottomrule[0.05em]
\end{tabular}}
\end{table*}

\begin{figure*}[!htbp]
	\centering
	\scriptsize
	\renewcommand{\h}{0.105}
	\renewcommand{\wa}{0.12}
	\newcommand{\wb}{0.16}
	\renewcommand{\g}{-0.7mm}
	\renewcommand{\tabcolsep}{1.8pt}
	\renewcommand{\arraystretch}{1}
        \resizebox{0.9\linewidth}{!} {
		\begin{tabular}{cc}			
			\renewcommand{\name}{figures/arad_929/ARAD_1K_0929_16_}
			\renewcommand{\h}{0.12}
			\renewcommand{\w}{0.2}
			\begin{tabular}{cc}
				\begin{adjustbox}{valign=t}
					\begin{tabular}{c}%
		         	\includegraphics[trim={74 40 0 40 },clip, width=0.275\textwidth]{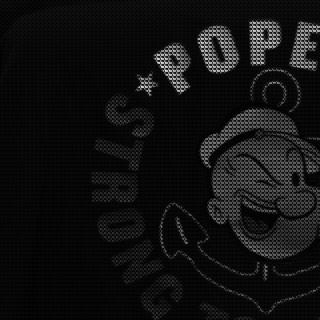}
						\\
						ARAD 929: Mosaic 
					\end{tabular}
				\end{adjustbox}
				\begin{adjustbox}{valign=t}
					\begin{tabular}{cccccc}
						\includegraphics[trim={125 215 115 40 },clip,height=\h \textwidth, width=\w \textwidth]{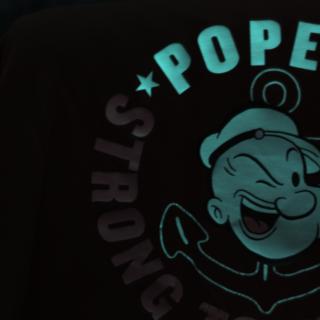} \hspace{\g} &
						\includegraphics[trim={125 215 115 40 },clip,height=\h \textwidth, width=\w \textwidth]{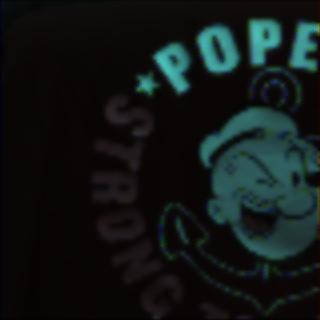} \hspace{\g} &
						\includegraphics[trim={125 215 115 40 },clip,height=\h \textwidth, width=\w \textwidth]{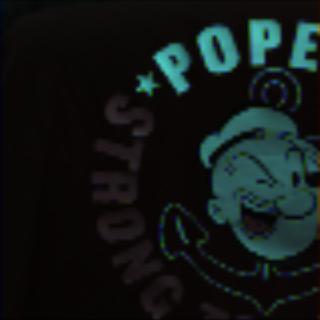} &
						\includegraphics[trim={125 215 115 40 },clip,height=\h \textwidth, width=\w \textwidth]{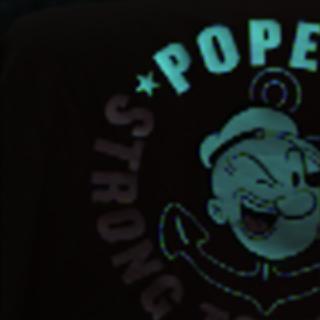} \hspace{\g} 
						\\
						GT  &
						BTES\cite{BTES} & WB\cite{WB} &
						PPID~\cite{PPID} 
						\\
						\vspace{-1.5mm}
						\\
						\includegraphics[trim={125 215 115 40 },clip,height=\h \textwidth, width=\w \textwidth]{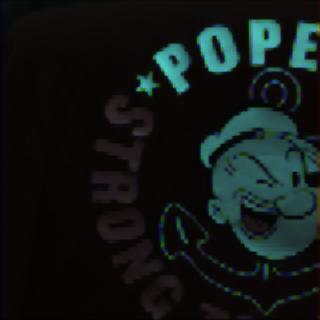} \hspace{\g} &
						\includegraphics[trim={125 215 115 40 },clip,height=\h \textwidth, width=\w \textwidth]{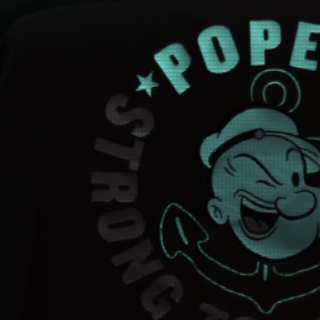} \hspace{\g} &
						\includegraphics[trim={125 215 115 40 },clip,height=\h \textwidth, width=\w \textwidth]{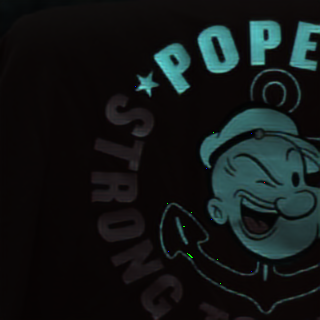}
						\hspace{\g} &		
						\includegraphics[trim={125 215 115 40},clip,height=\h \textwidth, width=\w \textwidth]{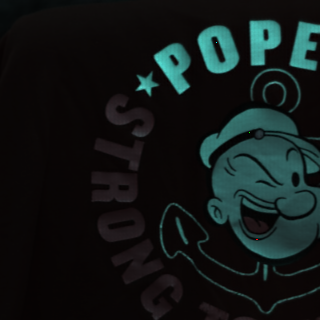} 
						\\ 
						GRMR\cite{GRMR}  \hspace{\g} &	MCAN \cite{MCAN}  \hspace{\g} & InNet~\cite{InNet}
						&
						\textbf{FDM-Net} (ours)
						\\
					\end{tabular}
				\end{adjustbox}
			\end{tabular}	
		\end{tabular}
	}

        \resizebox{0.9\linewidth}{!} {
		\begin{tabular}{cc}			
			\renewcommand{\name}{figures/arad_907/ARAD_1K_0907_16_}
			\renewcommand{\h}{0.12}
			\renewcommand{\w}{0.2}
			\begin{tabular}{cc}
				\begin{adjustbox}{valign=t}
					\begin{tabular}{c}%
		         	\includegraphics[trim={74 0 0 80 },clip, width=0.275\textwidth]{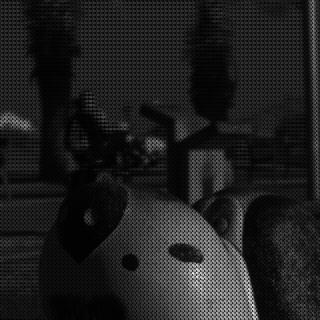}
						\\
						ARAD 907: Mosaic 
					\end{tabular}
				\end{adjustbox}
				\begin{adjustbox}{valign=t}
					\begin{tabular}{cccccc}
						\includegraphics[trim={160 45 40 90 },clip,height=\h \textwidth, width=\w \textwidth]{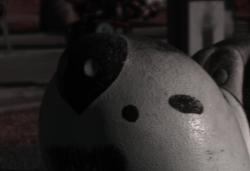} \hspace{\g} &
						\includegraphics[trim={160 45 40 90  },clip,height=\h \textwidth, width=\w \textwidth]{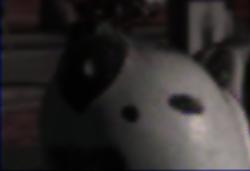} \hspace{\g} &
						\includegraphics[trim={160 45 40 90  },clip,height=\h \textwidth, width=\w \textwidth]{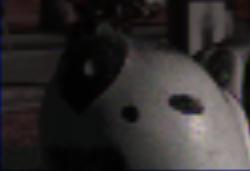} &
						\includegraphics[trim={160 45 40 90  },clip,height=\h \textwidth, width=\w \textwidth]{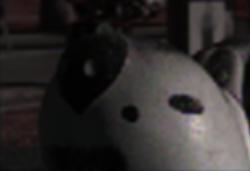} \hspace{\g} 
						\\
						GT  &
						BTES\cite{BTES} & WB\cite{WB} &
						PPID~\cite{PPID} 
						\\
						\vspace{-1.5mm}
						\\
						\includegraphics[trim={160 45 40 90  },clip,height=\h \textwidth, width=\w \textwidth]{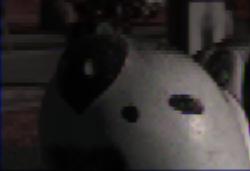} \hspace{\g} &
						\includegraphics[trim={160 45 40 90  },clip,height=\h \textwidth, width=\w \textwidth]{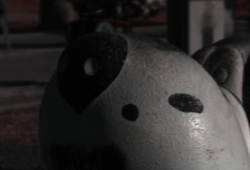} \hspace{\g} &
						\includegraphics[trim={160 45 40 90   },clip,height=\h \textwidth, width=\w \textwidth]{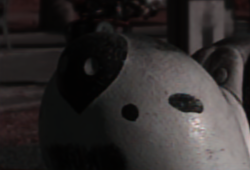}
						\hspace{\g} &		
						\includegraphics[trim={160 45 40 90    },clip,height=\h \textwidth, width=\w \textwidth]{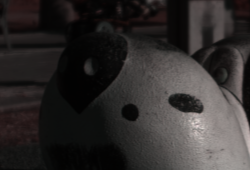} 
						\\ 
						GRMR\cite{GRMR}  \hspace{\g} &	MCAN \cite{MCAN}  \hspace{\g} & InNet~\cite{InNet}
						&
						\textbf{FDM-Net} (ours)
						\\
					\end{tabular}
				\end{adjustbox}
			\end{tabular}	
		\end{tabular}
	}
	\vspace{0.5mm}
	\caption{Visual comparison of \textbf{HSI demosaicing} methods (False color, R: 2, G: 11, B:16).} %
	  \vspace{-2mm}
	\label{fig_ARAD_907}
\end{figure*}

\section{DATASETS AND TRAINING}

\subsection{Datasets and Processing}



Attaining high accuracy in supervised MSFA demosaicing requires access to ground truth hyperspectral information for both training and evaluation purposes. We utilized the ARAD 1K dataset \cite{arad2022ntire,arad2022ntire_RGB2HSI} to meet this requirement, which comprises 384 images ranging from 400-1000nm and includes 16-channel hyperspectral images spanning a wide range of wavelengths. The ground truth hyperspectral information is presented as 480 x 512 spatial resolution images over the 16 spectral bands.
Previous methods for spectral demosaicing have employed diverse techniques, but they have often relied on limited datasets for evaluation, with some recent works testing on fewer than 10 images. In contrast, we validated our model on 50 HSI cubes, which features various scenes and also served as the validation dataset for the NTIRE 2022 spectral demosaicing challenge \cite{arad2022ntire,arad2022ntire_RGB2HSI}. 

\subsection{Training Details}

The proposed demosaicing model was trained from scratch using randomly initialized weights drawn from a normal distribution. We employed the Adam optimizer \cite{kingma2014adam} with a learning rate of 0.0001 that was halved every 1000 epochs. 
The weights of our loss are: $\alpha_1 = \alpha_2 = \alpha_3 = 1$.
For all compared methods and our FDM-Net, the patch size was set to $128 \times 128$, and the MSFA pattern was $4 \times 4$, resulting in 16 bands in the latent HSI. All networks were trained using one NVIDIA RTX 2080Ti-11GB GPU running Ubuntu 22.04.1. The FDM-Net was trained for 1600 epochs until convergence. In addition, for fair comparison, all the SOTA methods are retrained on ARAD1K \cite{arad2022ntire}.

\begin{figure*}[!htbp]
	\centering
	\scriptsize
	\renewcommand{\h}{0.105}
	\renewcommand{\wa}{0.12}
	\newcommand{\wb}{0.16}
	\renewcommand{\g}{-0.7mm}
	\renewcommand{\tabcolsep}{1.8pt}
	\renewcommand{\arraystretch}{1}
        \resizebox{0.9\linewidth}{!} {
		\begin{tabular}{cc}			
			\renewcommand{\name}{figures/arad_901/ARAD_1K_0901_16_}
			\renewcommand{\h}{0.12}
			\renewcommand{\w}{0.2}
			\begin{tabular}{cc}
				\begin{adjustbox}{valign=t}
					\begin{tabular}{c}%
		         	\includegraphics[trim={74 10 0 70 },clip, width=0.275\textwidth]{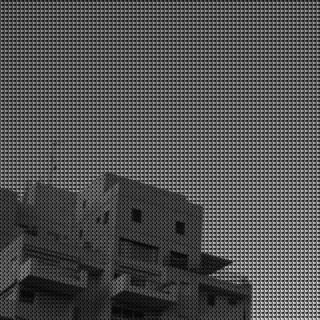}
						\\
						ARAD 901: Mosaic 
					\end{tabular}
				\end{adjustbox}
				\begin{adjustbox}{valign=t}
					\begin{tabular}{cccccc}
						\includegraphics[trim={110 30 150 260 },clip,height=\h \textwidth, width=\w \textwidth]{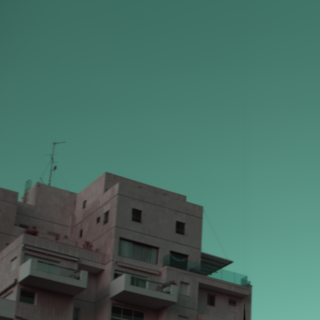} \hspace{\g} &
						\includegraphics[trim={110 30 150 260  },clip,height=\h \textwidth, width=\w \textwidth]{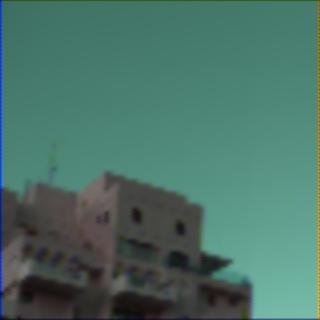} \hspace{\g} &
						\includegraphics[trim={110 30 150 260  },clip,height=\h \textwidth, width=\w \textwidth]{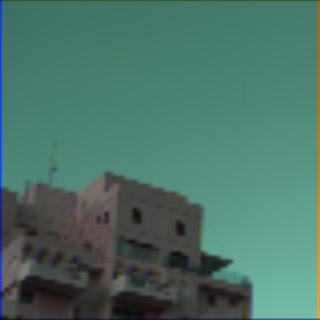} &
						\includegraphics[trim={110 30 150 260  },clip,height=\h \textwidth, width=\w \textwidth]{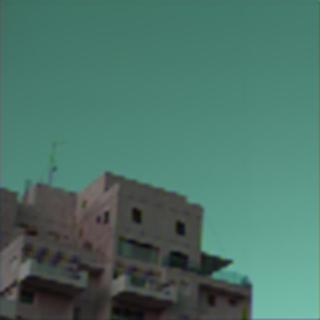} \hspace{\g} 
						\\
						GT  &
						BTES\cite{BTES} & WB\cite{WB} &
						PPID~\cite{PPID} 
						\\
						\vspace{-1.5mm}
						\\
						\includegraphics[trim={110 30 150 260  },clip,height=\h \textwidth, width=\w \textwidth]{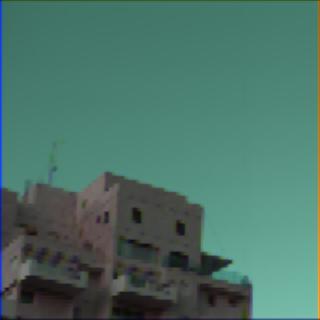} \hspace{\g} &
						\includegraphics[trim={110 30 150 260  },clip,height=\h \textwidth, width=\w \textwidth]{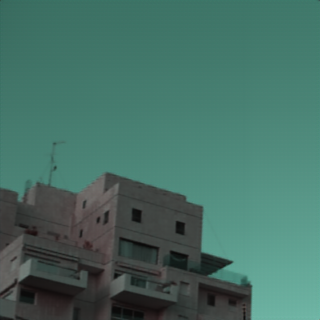} \hspace{\g} &
						\includegraphics[trim={110 30 150 260  },clip,height=\h \textwidth, width=\w \textwidth]{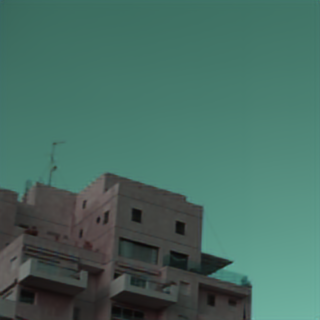}
						\hspace{\g} &		
						\includegraphics[trim={110 30 150 260  },clip,height=\h \textwidth, width=\w \textwidth]{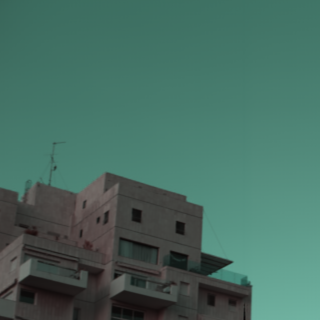} 
						\\ 
						GRMR\cite{GRMR}  \hspace{\g} & MCAN \cite{MCAN}  \hspace{\g} &	InNet~\cite{InNet}
						&
						\textbf{FDM-Net} (ours)
						\\
					\end{tabular}
				\end{adjustbox}
			\end{tabular}	
		\end{tabular}
	}
	\vspace{0.5mm}
	\caption{Visual comparison of \textbf{HSI demosaicing} methods (False color, R: 2, G: 11, B:16).} %
	  \vspace{-2mm}
	\label{fig_ARAD_901}
\end{figure*}

\begin{figure}[!htbp]
\centering
\includegraphics[width=3.3in]{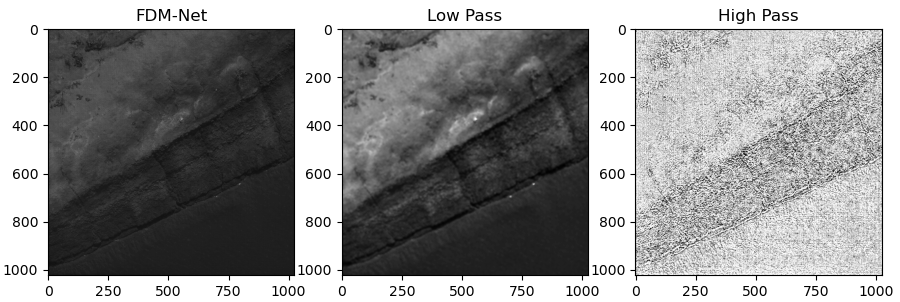}
\includegraphics[width=3.3in]{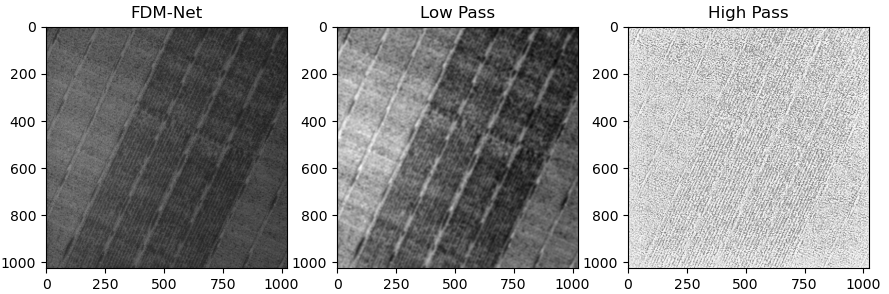}
\vspace{-4.8mm}
\caption{Illustration of the results on real data captured by our-self with IMEC $4 \times 4$ hyperspectral camera. }
\label{fig_imec}
\end{figure}

\begin{figure}[!htbp]
	\centering
	\scriptsize
	\renewcommand{\h}{0.105}
	\renewcommand{\wa}{0.12}
	\newcommand{\wb}{0.16}
	\renewcommand{\g}{-0.7mm}
	\renewcommand{\tabcolsep}{1.8pt}
	\renewcommand{\arraystretch}{1}
        \resizebox{1\linewidth}{!} {
		\begin{tabular}{c}	
                \normalsize
			\renewcommand{\name}{figures/ablation/ARAD_1K_0933_16_}
			\renewcommand{\h}{0.15}
			\renewcommand{\w}{0.09}
				\begin{adjustbox}{valign=t}
					\begin{tabular}{cccccc}
						\includegraphics[trim={130 50 85 140 },clip,height=\h \textwidth, width=\w \textwidth]{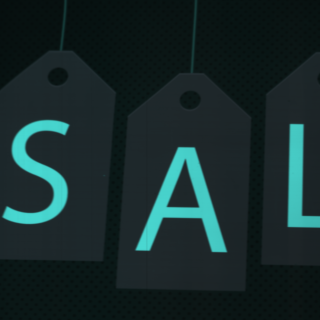}  &
						\includegraphics[trim={130 50 85 140 },clip,height=\h \textwidth, width=\w \textwidth]{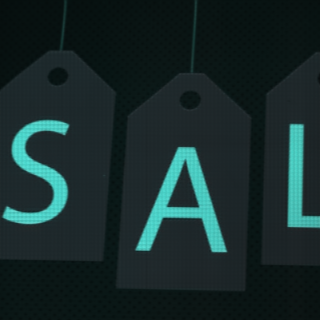}  &
						\includegraphics[trim={130 50 85 140 },clip,height=\h \textwidth, width=\w \textwidth]{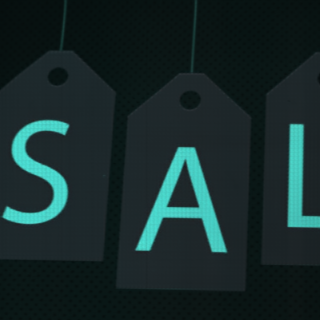}  &
					\includegraphics[trim={0 80 210 110 },clip,height=\h \textwidth, width=\w \textwidth]{\name gt.png}  &
						\includegraphics[trim={0 80 210 110 },clip,height=\h \textwidth, width=\w \textwidth]{\name our_blind.png}  &
						\includegraphics[trim={0 80 210 110 },clip,height=\h \textwidth, width=\w \textwidth]{\name our.png} 
                        \\ 
					GT  &	\textbf{W/O} & \textbf{W/} & GT  &	\textbf{W/O} & \textbf{W/}
						\\
					\end{tabular}
				\end{adjustbox}
			\end{tabular}
	}
	\vspace{-2mm}
	\caption{Visual comparison of \textbf{ablation studies}, \textbf{w/} means High+Low frequency based demosaicing using MaFormer, \textbf{w/o} is blind demosaicing using MaFormer. Zoom in for better view.} %
	  \vspace{-2mm}
	\label{fig_ablation}
\end{figure}

\section{EXPERIMENTAL RESULTS}

\subsection{Quantitative Experiments Metrics}
We quantitatively evaluate the performance of our proposed method by measuring the peak signal-to-noise ratio (PSNR), structure similarity (SSIM), Mean Relative Absolute Error (MRAE), and spectral angle mapper (SAM). PSNR and SSIM are conventional PQIs in image processing and computer vision that measure the similarity between the target and reference images based on mean squared error (MSE) and structural consistency, respectively. 
Finally, SAM is used to compute the spectral fidelity.

\subsection{Quantitative Comparison}
\begin{figure}[!htbp]
\centering
\includegraphics[width=1.6in]{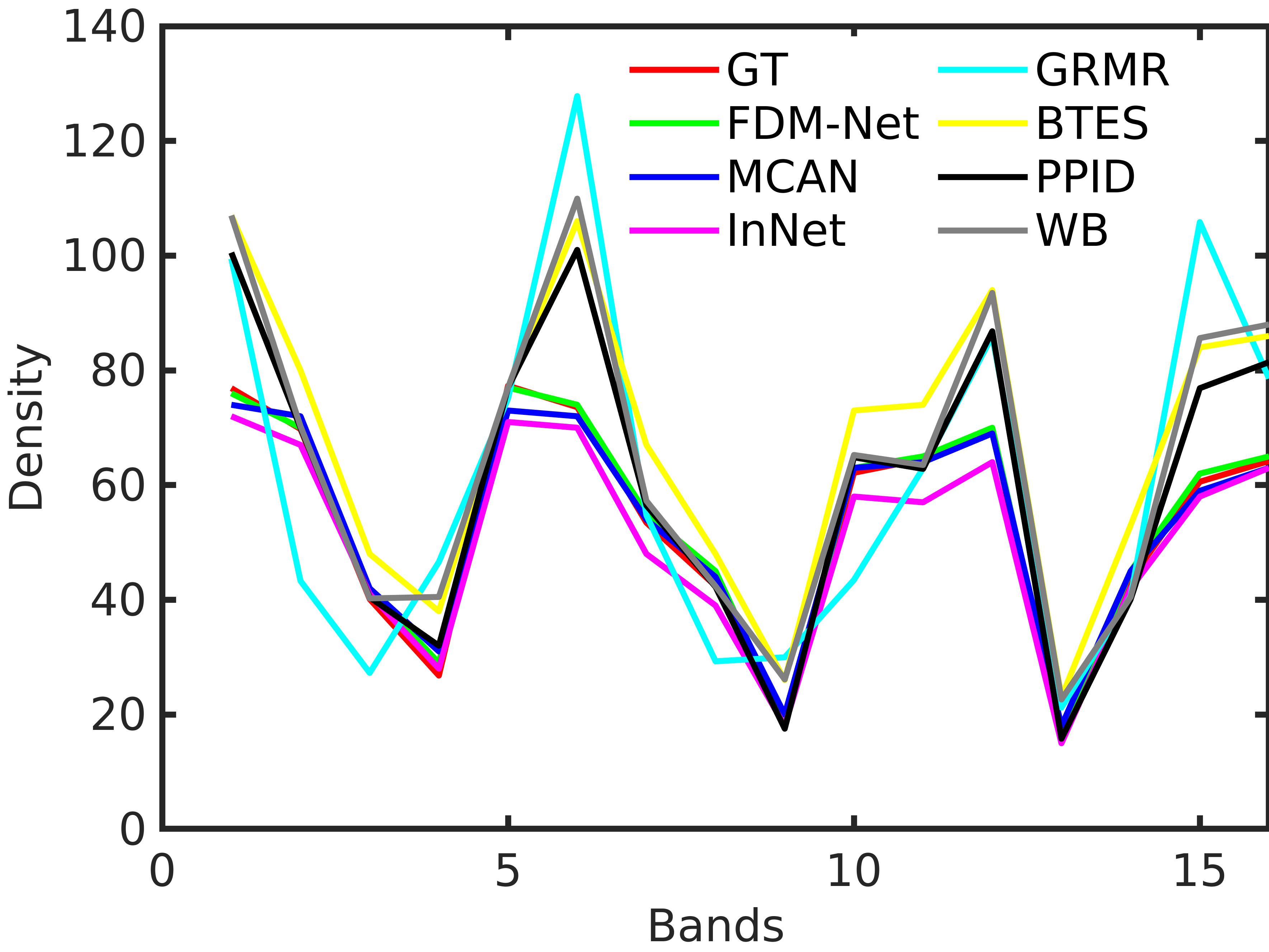}
\includegraphics[width=1.6in]{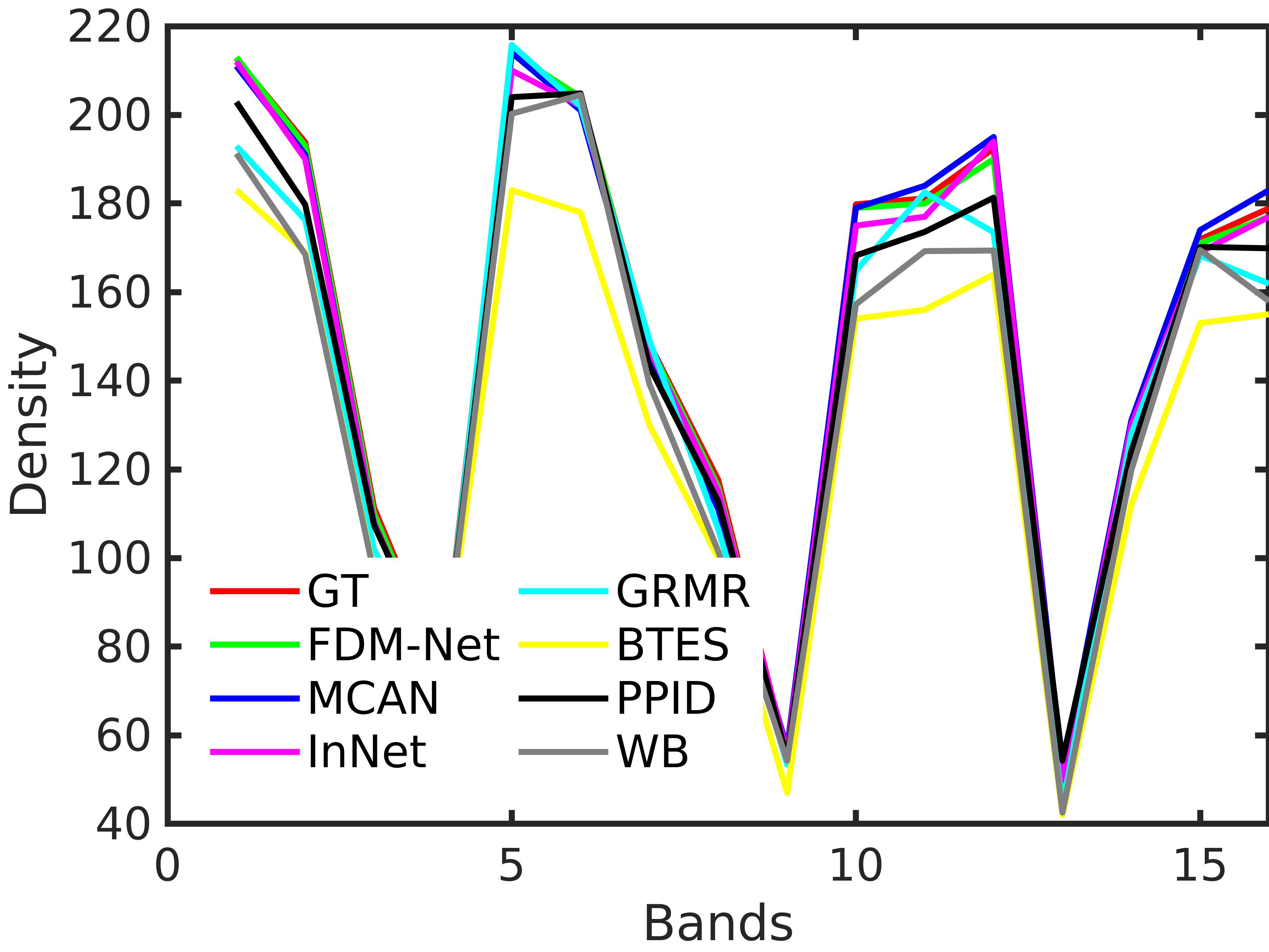}
\vspace{-1mm}
\caption{Illustration of the Spectral Density Curves on ARAD 907 HSI cube.  The curve for FDM-Net appears to be the most similar to the curve for GT, indicating that the spectral properties of FDM-Net are more closely aligned with the ground truth than the others.}
\label{fig_spectral_curve}
\end{figure}

We compare the proposed approach with the strong baselines: MCAN \cite{MCAN}, In-Net \cite{InNet}, PPID \cite{PPID}, GRMR \cite{GRMR}, and also classical WB \cite{WB}, BTES \cite{BTES}. 
The comparisons between FDM-Net, its light version: FDM-Net-L, and other SOTA methods are listed in Tab.~\ref{Table:PQI_1}. As can be observed: Our FDM-Net outperforms SOTA methods by a large margin. Specifically, FDM-Net surpasses the recent best algorithm MCAN by 6.01 dB, the classic PPID by 13.25dB, on the averaged results of 50 HSIs. 
In addition, our method exceeds all the counterparts with 7dB on three sub-datasets.
These results demonstrate the effectiveness of our method.
Furthermore, Tab.~\ref{Table:PQI_1} and Fig.~\ref{fig_spectral_curve} also present the spectral angle mapper (SAM) and curve of all tested methods. Our proposed FDM-Net demonstrates the highest correlation and coincidence with the reference, indicating its superiority in achieving spectral-dimension consistency reconstruction.

\subsection{Visual Comparison}
In this subsection, we present a comparison of our FDM-Net with other state-of-the-art (SOTA) methods for MSFA demosaicing, as depicted in Fig.~\ref{fig_ARAD_907}, \ref{fig_ARAD_901}. The left side of each image represents the observation data, which is the raw mosaicked image, while the right side displays zoomed-in patches of the red boxes in the entire hyperspectral images (HSIs).
Our FDM-Net exhibits superior performance compared to the other methods by producing visually pleasing HSIs with more detailed content, cleaner textures, and fewer artifacts, while simultaneously preserving the spatial smoothness of homogeneous regions. In contrast, previous methods have either yielded over-smooth results, which compromise fine-grained structures, or have introduced undesired chromatic artifacts and blotchy textures that are not present in the ground truth. Furthermore, Fig. \ref{fig_imec} demonstrates the efficacy of FDM-Net in processing real data captured with an IMEC HSI camera. More detailed results and discussion please refer to the \textbf{supplementary material}.
\subsection{Ablation Study on Pipeline Components}
\begin{figure}
	\centering
	\scriptsize
	\renewcommand{\h}{0.105}
	\renewcommand{\wa}{0.12}
	\newcommand{\wb}{0.16}
	\renewcommand{\g}{-0.7mm}
	\renewcommand{\tabcolsep}{1.8pt}
	\renewcommand{\arraystretch}{1}
        \resizebox{1\linewidth}{!} {
		\begin{tabular}{c}	
                \normalsize
			\renewcommand{\name}{figures/FDMNet_light/}
			\renewcommand{\h}{0.15}
			\renewcommand{\w}{0.09}
				\begin{adjustbox}{valign=t} 
					\begin{tabular}{cccccc}
						\includegraphics[trim={50 120 210 130 },clip,height=\h \textwidth, width=\w \textwidth]{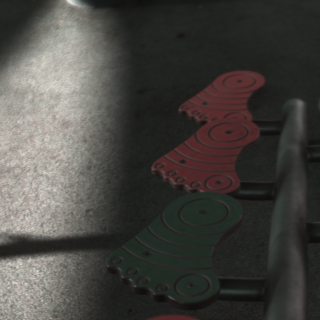}  &
						\includegraphics[trim={50 120 210 130 },clip,height=\h \textwidth, width=\w \textwidth]{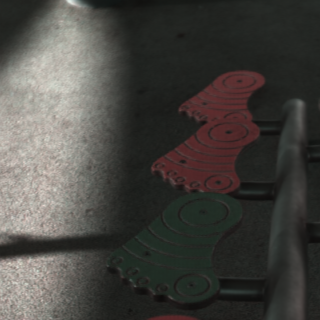}  &
						\includegraphics[trim={50 120 210 130 },clip,height=\h \textwidth, width=\w \textwidth]{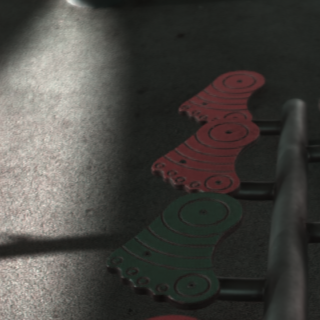}  &
					\includegraphics[trim={50 120 210 130 },clip,height=\h \textwidth, width=\w \textwidth]{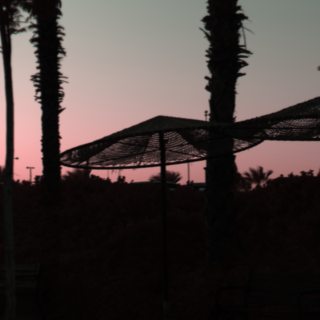}  &
						\includegraphics[trim={50 120 210 130 },clip,height=\h \textwidth, width=\w \textwidth]{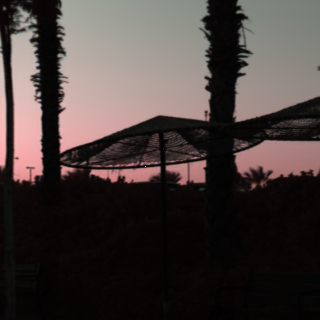}  &
						\includegraphics[trim={50 120 210 130 },clip,height=\h \textwidth, width=\w \textwidth]{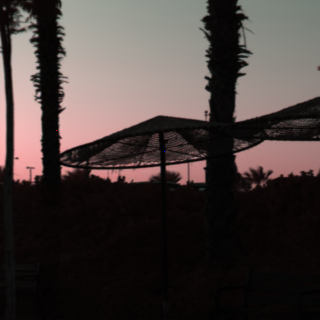} 
                        \\ 
					GT  & FDM-Net-L & FDM-Net & GT  &	FDM-Net-L & FDM-Net
						\\
					\end{tabular}
				\end{adjustbox}
			\end{tabular}
	}
	\vspace{-2mm}
	\caption{Visual comparison of \textbf{ablation studies}, FDM-Net in normal size with $N_i$=2, FDM-Net-L is a light one with $N_i$=1.} %
	  \vspace{-2mm}
	\label{fig_ablation_light}
\end{figure}

In order to examine the necessity of each component within our pipeline, we performed an ablation study on two distinct configurations of our method. These configurations included: 1) blind demosaicing using the MaFormer model, 2) utilizing the FDM-Net model with $N_i$ STMC blocks, with $N_i$ being fine-tuned for $i=1,2,3,4$, and then we evaluate the performance of these pipelines. The qualitative results of our ablation study are depicted in Fig.~\ref{fig_ablation} and \ref{fig_ablation_light}, while the quantitative results are presented in Tab.~\ref{Table:PQI_ablation} and \ref{Table:PQI_ablation_light}.
By comparing the blind demosaicing approach to the frequency-driven demosaicing method, we found that the latter demonstrated improvement in all metrics. This provides evidence that the method described in Sec.~\ref{model:overview}, which separates high and low frequency components, is superior to methods that do not. Additionally, we observed that increasing the value of $N_i$ from 1 to 2 resulted in a performance improvement of 1.59dB, but it also increased the running time by 1.86 times.

\begin{table}
\centering
\renewcommand{\arraystretch}{1.3}%
\caption{The PSNR, SSIM, SAM(lower is better), MRAE(lower is better) scores for the \textbf{ablation studies} of frequency.}
\label{Table:PQI_ablation}
\scalebox{0.67}{
\begin{tabular}{ccccccc}
\toprule[0.05em]
High& Low &  \multirow{2}{*}{Method} &	\multirow{2}{*}{PSNR $\uparrow$} & \multirow{2}{*}{SSIM $\uparrow$}  & \multirow{2}{*}{SAM $\downarrow$} & \multirow{2}{*}{MRAE $\downarrow$}\\
Frequency &  Frequency &  &  &  &  &\\
\Xcline{1-7}{0.05pt}
\xmark  & \xmark & MaFormer  & 47.64 & 0.994 & 0.014 & 0.019\\
\cmark  & \cmark & FDM-Net  & \textbf{49.23}  & \textbf{0.996} & \textbf{0.013} & \textbf{0.017}\\
\Xcline{1-7}{0.05pt}
\vspace{-1mm}
\end{tabular}}
\end{table}

\begin{table}
\centering
\renewcommand{\arraystretch}{1.3}%
\caption{The PSNR, SSIM, SAM, MRAE of the \textbf{ablation studies} of model size for the proposed FDM-Net, i.e., FDM-Net with normal size, $N_i=2, i-1,2,3,4$, and light FDM-Net with $N_i=1$.}
\label{Table:PQI_ablation_light}
\scalebox{0.68}{
\begin{tabular}{ccccccc}
\toprule[0.05em]
Model Size & STMC Block &	PSNR $\uparrow$ & SSIM $\uparrow$  & SAM $\downarrow$ & MRAE $\downarrow$ & Time(s)\\
\midrule
Light & $N_i$=1  & 48.60 & 0.995 & 0.014 & 0.018 & \textbf{0.029}\\
Normal & $N_i$=2 & \textbf{49.23}  & \textbf{0.996} & \textbf{0.013} & \textbf{0.017} & 0.053\\
\bottomrule
\vspace{-1mm}
\end{tabular}}
\end{table}

\section{Conclusion}

This paper has proposed a novel HSI demosaicing method that is driven by both high and low frequencies. The proposed method leverages Fourier zero-padding to quickly reconstruct the easy low frequency part, while a customized transformer architecture effectively handles the challenging task of high pass HSI demosaicing. By introducing a joint spatial and frequency loss, our approach enhances high frequency modeling while ensuring stable low frequency reconstruction.
Extensive evaluations of the proposed approach on a large testing dataset comprising 50 HSI cubes demonstrate that it achieves SOTA performance, outperforming SOTA 6.01dB.
Overall, the results indicate that focusing on the hard high frequency components has the potential to improve the accuracy and reliability of HSI demosaicing in various applications. 
Further research could explore the possibility of incorporating additional prior.

\definecolor{Gray}{gray}{0.90}
\begin{table*}[!htbp]
	\centering
	\setlength{\tabcolsep}{2.4pt}
	\renewcommand{\arraystretch}{1.2}%
	\caption{\textbf{Detailed Demosaicing results compared with other methods. Our FDM-Net significantly surpass other competitors}}
	\vspace{1mm}
	\label{Table:PQI_1}
	\scalebox{0.85}{
		\begin{tabular}{|c|>{\bfseries}c>{\bfseries}c>{\bfseries}c>{\bfseries}c|cccc|cccc|}
			\toprule[0.05em]
			\rowcolor{Gray}
			& \multicolumn{4}{c|}{\textbf{FDM-Net (Ours)}} & \multicolumn{4}{c|}{\textbf{MCAN}} & \multicolumn{4}{c|}{\textbf{InNet}} \\
			\rowcolor{Gray}
			HSIs	&	PSNR	&	SSIM	&	SAM	&	MRAE	&	\textbf{PSNR}	&	\textbf{SSIM}	&	\textbf{SAM}	&	\textbf{MRAE}	&	\textbf{PSNR}	&	\textbf{SSIM}	&	\textbf{SAM}	&	\textbf{MRAE}	\\	
			\midrule
			ARAD1K090116 	&	51.648	&	0.997	&	0.004	&	0.005	&	41.596	&	0.988	&	0.020	&	0.022	&	44.983	&	0.991	&	0.011	&	0.014	\\
			ARAD1K090216 	&	48.082	&	0.997	&	0.024	&	0.028	&	42.850	&	0.991	&	0.045	&	0.053	&	42.208	&	0.987	&	0.061	&	0.072	\\
			ARAD1K090316 	&	48.629	&	0.993	&	0.011	&	0.012	&	36.920	&	0.936	&	0.064	&	0.071	&	41.094	&	0.985	&	0.041	&	0.037	\\
			ARAD1K090416 	&	47.682	&	0.997	&	0.015	&	0.018	&	42.604	&	0.990	&	0.040	&	0.044	&	42.109	&	0.986	&	0.039	&	0.046	\\
			ARAD1K090516 	&	43.056	&	0.984	&	0.015	&	0.017	&	37.452	&	0.954	&	0.040	&	0.042	&	37.940	&	0.950	&	0.032	&	0.036	\\
			ARAD1K090616 	&	47.648	&	0.993	&	0.009	&	0.012	&	43.079	&	0.986	&	0.020	&	0.023	&	41.708	&	0.972	&	0.020	&	0.027	\\
			ARAD1K090716 	&	50.842	&	0.997	&	0.014	&	0.017	&	45.293	&	0.991	&	0.030	&	0.038	&	43.498	&	0.984	&	0.033	&	0.043	\\
			ARAD1K090816 	&	45.274	&	0.995	&	0.029	&	0.036	&	40.079	&	0.988	&	0.048	&	0.064	&	39.914	&	0.972	&	0.060	&	0.078	\\
			ARAD1K090916 	&	49.755	&	0.996	&	0.008	&	0.011	&	43.756	&	0.990	&	0.020	&	0.026	&	41.926	&	0.977	&	0.026	&	0.034	\\
			ARAD1K091016 	&	49.294	&	0.996	&	0.011	&	0.015	&	44.458	&	0.990	&	0.026	&	0.034	&	43.912	&	0.987	&	0.027	&	0.036	\\
			ARAD1K091116 	&	53.020	&	0.998	&	0.010	&	0.012	&	46.716	&	0.992	&	0.026	&	0.029	&	45.638	&	0.987	&	0.027	&	0.034	\\
			ARAD1K091216 	&	39.662	&	0.993	&	0.024	&	0.027	&	35.010	&	0.975	&	0.041	&	0.053	&	33.140	&	0.944	&	0.055	&	0.074	\\
			ARAD1K091316 	&	48.726	&	0.996	&	0.008	&	0.009	&	41.912	&	0.988	&	0.021	&	0.026	&	41.695	&	0.983	&	0.019	&	0.023	\\
			ARAD1K091416 	&	49.943	&	0.995	&	0.016	&	0.019	&	46.012	&	0.989	&	0.035	&	0.043	&	44.501	&	0.984	&	0.035	&	0.044	\\
			ARAD1K091516 	&	47.226	&	0.995	&	0.015	&	0.020	&	43.193	&	0.987	&	0.031	&	0.039	&	40.915	&	0.975	&	0.038	&	0.049	\\
			ARAD1K091616 	&	53.799	&	0.998	&	0.015	&	0.018	&	46.397	&	0.993	&	0.042	&	0.047	&	47.605	&	0.989	&	0.037	&	0.044	\\
			ARAD1K091716 	&	48.404	&	0.996	&	0.007	&	0.008	&	40.293	&	0.986	&	0.025	&	0.028	&	42.113	&	0.988	&	0.017	&	0.020	\\
			ARAD1K091816 	&	53.342	&	0.997	&	0.004	&	0.005	&	45.133	&	0.989	&	0.016	&	0.024	&	48.285	&	0.992	&	0.009	&	0.012	\\
			ARAD1K091916 	&	48.149	&	0.998	&	0.008	&	0.009	&	43.597	&	0.994	&	0.020	&	0.023	&	42.299	&	0.992	&	0.018	&	0.021	\\
			ARAD1K092016 	&	43.598	&	0.989	&	0.028	&	0.037	&	40.435	&	0.982	&	0.054	&	0.065	&	39.952	&	0.970	&	0.057	&	0.079	\\
			ARAD1K092116 	&	53.018	&	0.997	&	0.012	&	0.015	&	47.913	&	0.993	&	0.029	&	0.038	&	46.338	&	0.987	&	0.032	&	0.041	\\
			ARAD1K092216 	&	47.805	&	0.997	&	0.009	&	0.010	&	42.061	&	0.992	&	0.023	&	0.028	&	41.129	&	0.986	&	0.025	&	0.029	\\
			ARAD1K092316 	&	45.168	&	0.992	&	0.017	&	0.022	&	40.907	&	0.982	&	0.032	&	0.041	&	39.132	&	0.968	&	0.038	&	0.052	\\
			ARAD1K092416 	&	49.216	&	0.996	&	0.009	&	0.011	&	42.483	&	0.988	&	0.021	&	0.028	&	42.557	&	0.982	&	0.022	&	0.028	\\
			ARAD1K092516 	&	50.128	&	0.997	&	0.009	&	0.012	&	45.396	&	0.994	&	0.022	&	0.029	&	45.149	&	0.992	&	0.019	&	0.025	\\
			ARAD1K092616 	&	50.816	&	0.996	&	0.023	&	0.032	&	47.260	&	0.990	&	0.049	&	0.069	&	44.727	&	0.981	&	0.063	&	0.086	\\
			ARAD1K092716 	&	48.117	&	0.996	&	0.013	&	0.014	&	41.822	&	0.987	&	0.031	&	0.035	&	41.693	&	0.986	&	0.029	&	0.033	\\
			ARAD1K092816 	&	41.490	&	0.985	&	0.044	&	0.052	&	38.339	&	0.966	&	0.066	&	0.086	&	36.442	&	0.938	&	0.082	&	0.113	\\
			ARAD1K092916 	&	46.020	&	0.992	&	0.030	&	0.037	&	39.259	&	0.974	&	0.082	&	0.101	&	40.664	&	0.957	&	0.064	&	0.076	\\
			ARAD1K093016 	&	48.675	&	0.996	&	0.010	&	0.013	&	42.521	&	0.990	&	0.024	&	0.034	&	43.001	&	0.987	&	0.022	&	0.028	\\
			ARAD1K093116 	&	51.699	&	0.997	&	0.004	&	0.007	&	45.832	&	0.993	&	0.017	&	0.024	&	46.247	&	0.991	&	0.011	&	0.016	\\
			ARAD1K093216 	&	45.773	&	0.993	&	0.009	&	0.011	&	35.575	&	0.958	&	0.054	&	0.081	&	40.879	&	0.987	&	0.025	&	0.027	\\
			ARAD1K093316 	&	43.582	&	0.991	&	0.017	&	0.022	&	37.880	&	0.965	&	0.066	&	0.084	&	39.716	&	0.977	&	0.050	&	0.056	\\
			ARAD1K093416 	&	51.068	&	0.997	&	0.010	&	0.016	&	47.002	&	0.991	&	0.026	&	0.063	&	45.602	&	0.990	&	0.025	&	0.042	\\
			ARAD1K093516 	&	44.234	&	0.991	&	0.016	&	0.021	&	40.954	&	0.983	&	0.028	&	0.037	&	38.755	&	0.969	&	0.033	&	0.045	\\
			ARAD1K093616 	&	44.797	&	0.994	&	0.013	&	0.018	&	40.823	&	0.986	&	0.027	&	0.041	&	40.041	&	0.980	&	0.028	&	0.038	\\
			ARAD1K093716 	&	52.013	&	0.997	&	0.014	&	0.021	&	47.991	&	0.993	&	0.037	&	0.059	&	46.512	&	0.992	&	0.034	&	0.050	\\
			ARAD1K093816 	&	43.117	&	0.992	&	0.017	&	0.023	&	39.648	&	0.982	&	0.032	&	0.048	&	38.722	&	0.974	&	0.036	&	0.049	\\
			ARAD1K093916 	&	52.190	&	0.996	&	0.013	&	0.017	&	48.554	&	0.993	&	0.028	&	0.035	&	47.207	&	0.989	&	0.029	&	0.040	\\
			ARAD1K094016 	&	44.691	&	0.994	&	0.016	&	0.021	&	39.697	&	0.986	&	0.031	&	0.044	&	39.901	&	0.982	&	0.034	&	0.046	\\
			ARAD1K094116 	&	48.969	&	0.995	&	0.018	&	0.024	&	45.662	&	0.991	&	0.032	&	0.046	&	43.960	&	0.986	&	0.039	&	0.052	\\
			ARAD1K094216 	&	47.487	&	0.996	&	0.008	&	0.012	&	42.336	&	0.990	&	0.020	&	0.033	&	41.663	&	0.985	&	0.019	&	0.027	\\
			ARAD1K094316 	&	52.551	&	0.998	&	0.006	&	0.007	&	43.046	&	0.994	&	0.018	&	0.021	&	47.079	&	0.994	&	0.016	&	0.019	\\
			ARAD1K094416 	&	48.019	&	0.996	&	0.015	&	0.019	&	43.765	&	0.990	&	0.037	&	0.044	&	43.480	&	0.986	&	0.041	&	0.051	\\
			ARAD1K094516 	&	52.938	&	0.998	&	0.011	&	0.014	&	48.726	&	0.995	&	0.030	&	0.038	&	47.715	&	0.992	&	0.031	&	0.039	\\
			ARAD1K094616 	&	51.571	&	0.997	&	0.013	&	0.016	&	47.802	&	0.994	&	0.029	&	0.033	&	45.627	&	0.990	&	0.030	&	0.038	\\
			ARAD1K094716 	&	57.270	&	0.998	&	0.016	&	0.021	&	53.460	&	0.996	&	0.042	&	0.050	&	51.726	&	0.993	&	0.046	&	0.062	\\
			ARAD1K094816 	&	52.415	&	0.998	&	0.006	&	0.009	&	47.046	&	0.994	&	0.020	&	0.025	&	46.392	&	0.992	&	0.015	&	0.022	\\
			ARAD1K094916 	&	45.846	&	0.994	&	0.018	&	0.023	&	42.039	&	0.987	&	0.036	&	0.048	&	40.264	&	0.978	&	0.044	&	0.058	\\
			ARAD1K095016 	&	51.319	&	0.019	&	0.02	&	0.008	&	48.538	&	0.992	&	0.036	&	0.047	&	46.117	&	0.986	&	0.044	&	0.057	\\
			\hline
	\end{tabular}}
\end{table*}

\begin{figure*}[!htbp]
	\centering
	\scriptsize
	\renewcommand{\h}{0.105}
	\renewcommand{\wa}{0.12}
	\newcommand{\wb}{0.16}
	\renewcommand{\g}{-0.7mm}
	\renewcommand{\tabcolsep}{1.8pt}
	\renewcommand{\arraystretch}{1}
	\resizebox{0.85\linewidth}{!} {
		\begin{tabular}{cc}			
			\renewcommand{\name}{figures/arad_905/ARAD_1K_0905_16_}
			\renewcommand{\h}{0.12}
			\renewcommand{\w}{0.2}
			\begin{tabular}{cc}
				\begin{adjustbox}{valign=t}
					\begin{tabular}{cccccc}
						\includegraphics[trim={175 185 55 75 },clip,height=\h \textwidth, width=\w \textwidth]{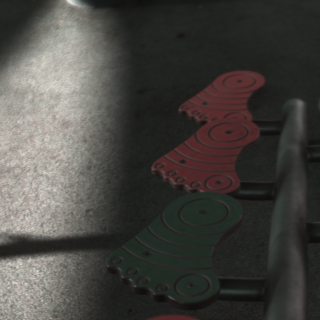} \hspace{\g} &
						\includegraphics[trim={175 185 55 75 },clip,height=\h \textwidth, width=\w \textwidth]{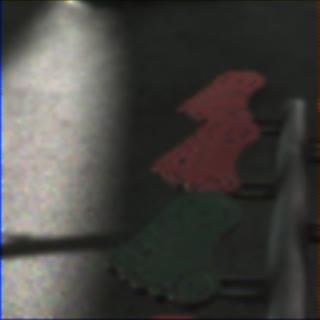} \hspace{\g} &
						\includegraphics[trim={175 185 55 75 },clip,height=\h \textwidth, width=\w \textwidth]{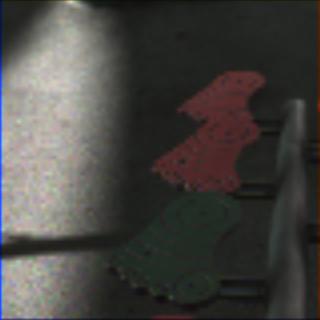} &
						\includegraphics[trim={175 185 55 75 },clip,height=\h \textwidth, width=\w \textwidth]{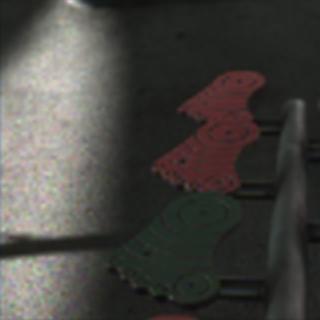} \hspace{\g} 
						\\
						GT &
						BTES & WB &
						PPID 
						\\
						\includegraphics[trim={175 185 55 75 },clip,height=\h \textwidth, width=\w \textwidth]{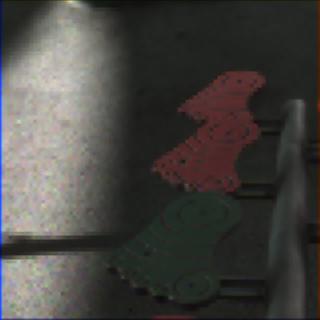} \hspace{\g} &
						\includegraphics[trim={175 185 55 75 },clip,height=\h \textwidth, width=\w \textwidth]{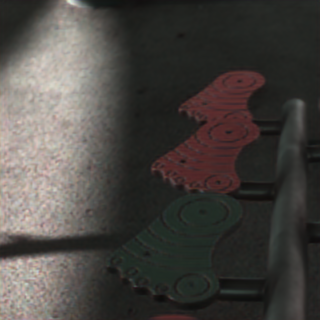} \hspace{\g} &
						\includegraphics[trim={175 185 55 75 },clip,height=\h \textwidth, width=\w \textwidth]{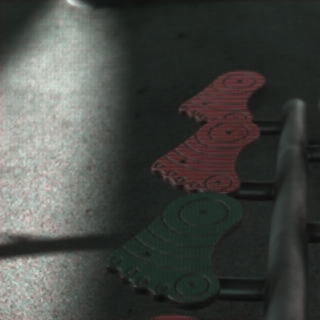}
						\hspace{\g} &		
						\includegraphics[trim={175 185 55 75 },clip,height=\h \textwidth, width=\w \textwidth]{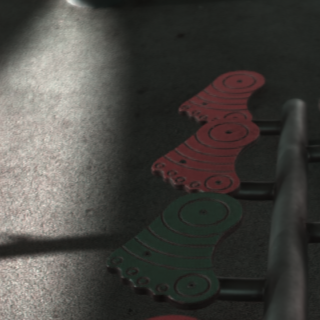} 
						\\ 
						GRMR  \hspace{\g} &	InNet  \hspace{\g} & MCAN
						&
						\textbf{FDM-Net} (ours)
						\\
					\end{tabular}
				\end{adjustbox}
			\end{tabular}	
		\end{tabular}
	}
	\resizebox{0.85\linewidth}{!} {
		\begin{tabular}{cc}			
			\renewcommand{\name}{figures/arad_903/ARAD_1K_0903_16_}
			\renewcommand{\h}{0.12}
			\renewcommand{\w}{0.2}
			\begin{tabular}{cc}
				\begin{adjustbox}{valign=t}
					\begin{tabular}{cccccc}
						\includegraphics[trim={265 45 5 235 },clip,height=\h \textwidth, width=\w \textwidth]{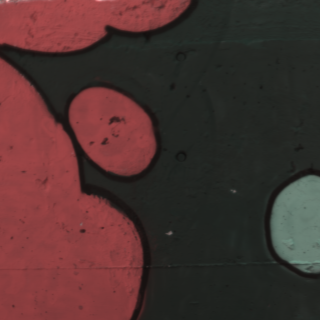} \hspace{\g} &
						\includegraphics[trim={265 45 5 235 },clip,height=\h \textwidth, width=\w \textwidth]{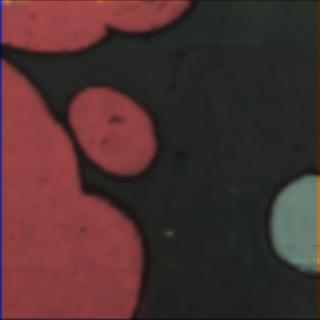} \hspace{\g} &
						\includegraphics[trim={265 45 5 235 },clip,height=\h \textwidth, width=\w \textwidth]{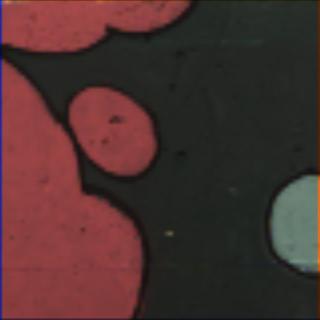} &
						\includegraphics[trim={265 45 5 235 },clip,height=\h \textwidth, width=\w \textwidth]{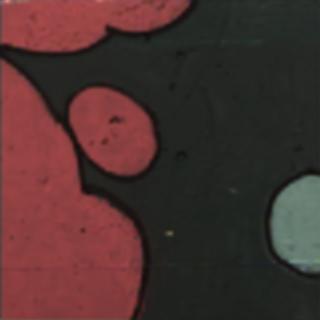} \hspace{\g} 
						\\
						GT &
						BTES & WB &
						PPID 
						\\
						\vspace{-2mm}
						\\
						\includegraphics[trim={265 45 5 235 },clip,height=\h \textwidth, width=\w \textwidth]{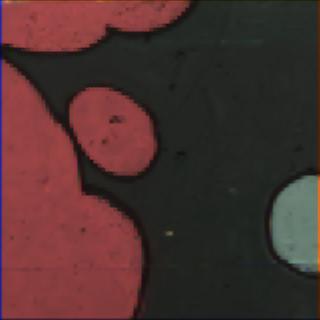} \hspace{\g} &
						\includegraphics[trim={265 45 5 235 },clip,height=\h \textwidth, width=\w \textwidth]{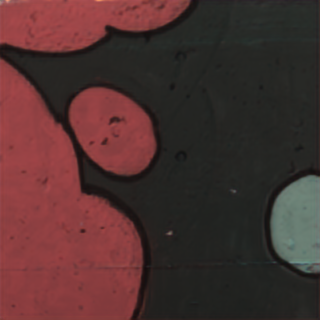} \hspace{\g} &
						\includegraphics[trim={265 45 5 235 },clip,height=\h \textwidth, width=\w \textwidth]{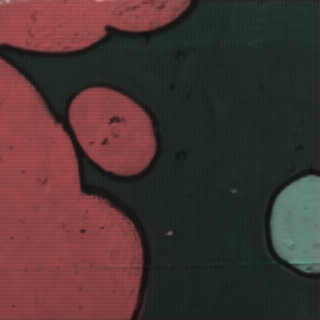}
						\hspace{\g} &		
						\includegraphics[trim={265 45 5 235 },clip,height=\h \textwidth, width=\w \textwidth]{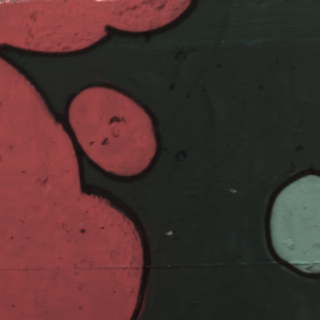} 
						\\ 
						GRMR  \hspace{\g} &	InNet \hspace{\g} & MCAN
						&
						\textbf{FDM-Net} (ours)
						\\
					\end{tabular}
				\end{adjustbox}
			\end{tabular}	
		\end{tabular}
	}
	\resizebox{0.85\linewidth}{!} {
		\begin{tabular}{cc}			
			\renewcommand{\name}{figures/arad_911/ARAD_1K_0911_16_}
			\renewcommand{\h}{0.12}
			\renewcommand{\w}{0.2}
			\begin{tabular}{cc}
				\begin{adjustbox}{valign=t}
					\begin{tabular}{cccccc}
						\includegraphics[trim={65 45 205 235 },clip,height=\h \textwidth, width=\w \textwidth]{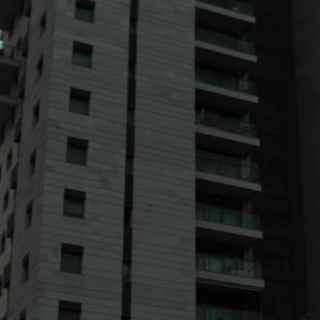} \hspace{\g} &
						\includegraphics[trim={65 45 205 235 },clip,height=\h \textwidth, width=\w \textwidth]{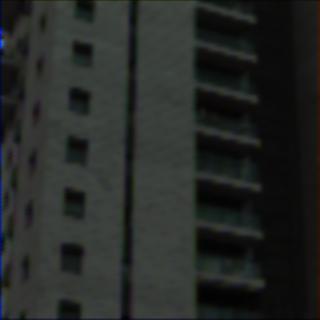} \hspace{\g} &
						\includegraphics[trim={65 45 205 235 },clip,height=\h \textwidth, width=\w \textwidth]{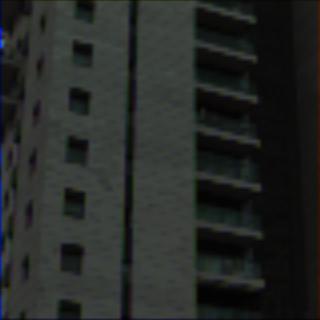} &
						\includegraphics[trim={65 45 205 235 },clip,height=\h \textwidth, width=\w \textwidth]{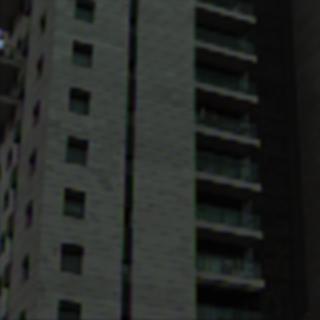} \hspace{\g} 
						\\
						GT &
						BTES & WB &
						PPID 
						\\
						\vspace{-2mm}
						\\
						\includegraphics[trim={65 45 205 235 },clip,height=\h \textwidth, width=\w \textwidth]{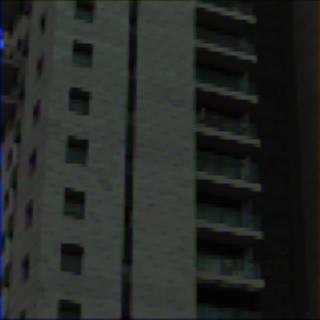} \hspace{\g} &
						\includegraphics[trim={65 45 205 235 },clip,height=\h \textwidth, width=\w \textwidth]{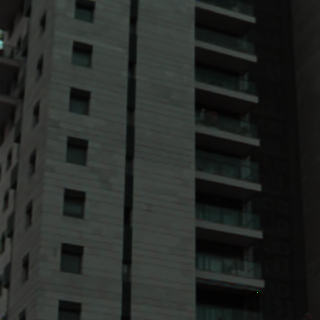} \hspace{\g} &
						\includegraphics[trim={65 45 205 235 },clip,height=\h \textwidth, width=\w \textwidth]{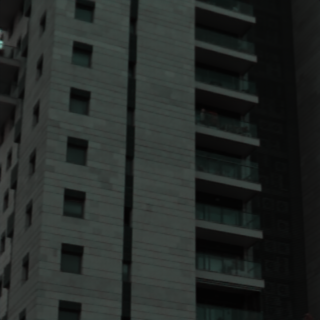}
						\hspace{\g} &		
						\includegraphics[trim={65 45 205 235 },clip,height=\h \textwidth, width=\w \textwidth]{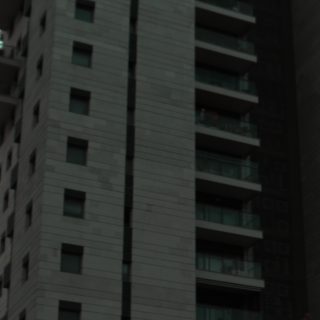} 
						\\ 
						GRMR  \hspace{\g} &	InNet \hspace{\g} & MCAN
						&
						\textbf{FDM-Net} (ours)
						\\
					\end{tabular}
				\end{adjustbox}
			\end{tabular}	
		\end{tabular}
	}
	\resizebox{0.85\linewidth}{!} {
		\begin{tabular}{cc}			
			\renewcommand{\name}{figures/arad_917/ARAD_1K_0917_16_}
			\renewcommand{\h}{0.12}
			\renewcommand{\w}{0.2}
			\begin{tabular}{cc}
				\begin{adjustbox}{valign=t}
					\begin{tabular}{cccccc}
						\includegraphics[trim={75 200 155 90 },clip,height=\h \textwidth, width=\w \textwidth]{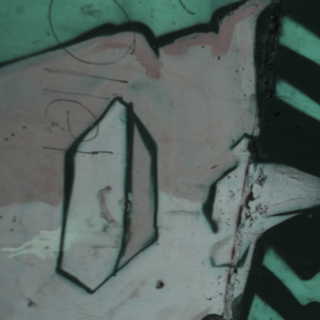} \hspace{\g} &
						\includegraphics[trim={75 200 155 90 },clip,height=\h \textwidth, width=\w \textwidth]{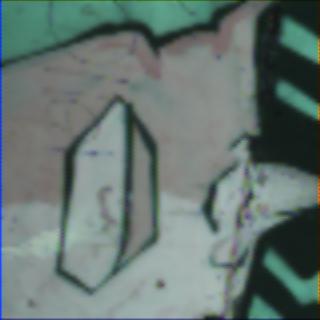} \hspace{\g} &
						\includegraphics[trim={75 200 155 90 },clip,height=\h \textwidth, width=\w \textwidth]{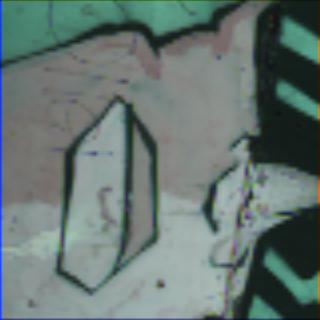} &
						\includegraphics[trim={75 200 155 90 },clip,height=\h \textwidth, width=\w \textwidth]{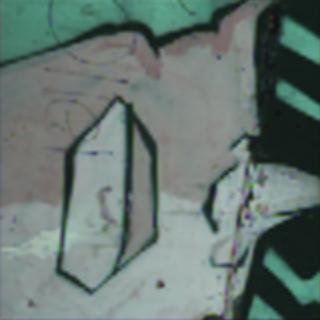} \hspace{\g} 
						\\
						GT &
						BTES & WB &
						PPID 
						\\
						\vspace{-2mm}
						\\
						\includegraphics[trim={75 200 155 90 },clip,height=\h \textwidth, width=\w \textwidth]{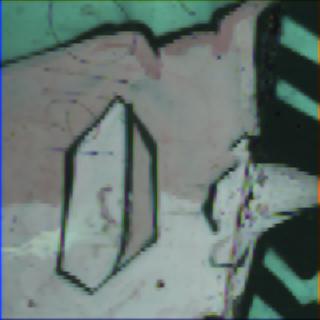} \hspace{\g} &
						\includegraphics[trim={75 200 155 90 },clip,height=\h \textwidth, width=\w \textwidth]{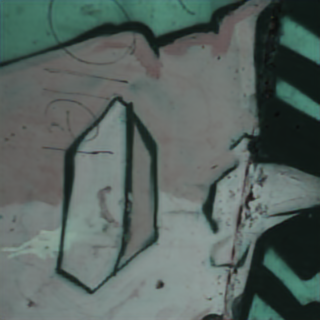} \hspace{\g} &
						\includegraphics[trim={75 200 155 90 },clip,height=\h \textwidth, width=\w \textwidth]{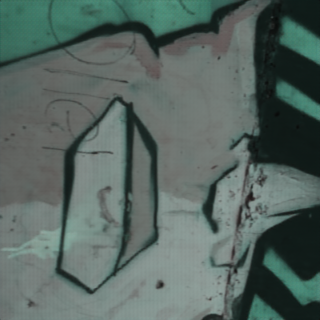}
						\hspace{\g} &		
						\includegraphics[trim={75 200 155 90 },clip,height=\h \textwidth, width=\w \textwidth]{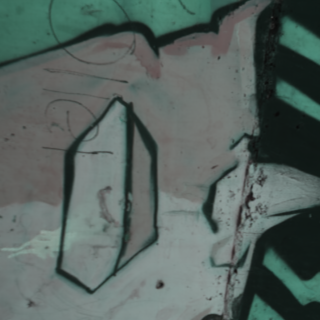} 
						\\ 
						GRMR  \hspace{\g} &	InNet  \hspace{\g} & MCAN
						&
						\textbf{FDM-Net} (ours)
						\\
					\end{tabular}
				\end{adjustbox}
			\end{tabular}	
		\end{tabular}
	}
	\resizebox{0.85\linewidth}{!} {
		\begin{tabular}{cc}			
			\renewcommand{\name}{figures/arad_940/ARAD_1K_0940_16_}
			\renewcommand{\h}{0.12}
			\renewcommand{\w}{0.2}
			\begin{tabular}{cc}
				\begin{adjustbox}{valign=t}
					\begin{tabular}{cccccc}
						\includegraphics[trim={105 185 105 75 },clip,height=\h \textwidth, width=\w \textwidth]{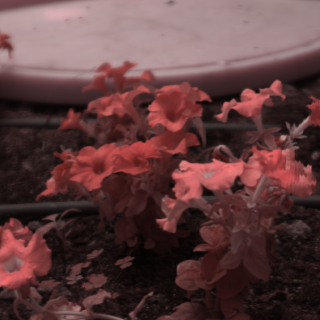} \hspace{\g} &
						\includegraphics[trim={105 185 105 75 },clip,height=\h \textwidth, width=\w \textwidth]{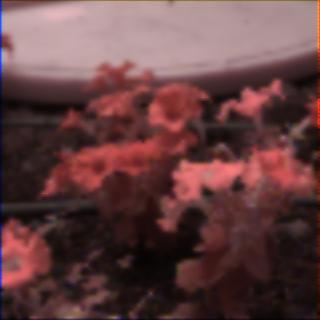} \hspace{\g} &
						\includegraphics[trim={105 185 105 75 },clip,height=\h \textwidth, width=\w \textwidth]{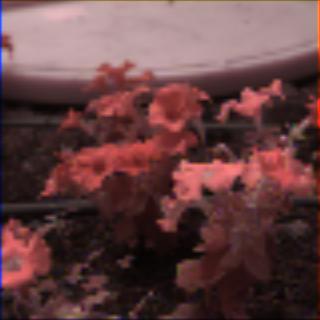} &
						\includegraphics[trim={105 185 105 75 },clip,height=\h \textwidth, width=\w \textwidth]{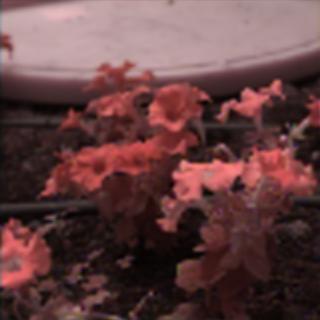} \hspace{\g} 
						\\
						GT &
						BTES & WB &
						PPID 
						\\
						\vspace{-2mm}
						\\
						\includegraphics[trim={105 185 105 75 },clip,height=\h \textwidth, width=\w \textwidth]{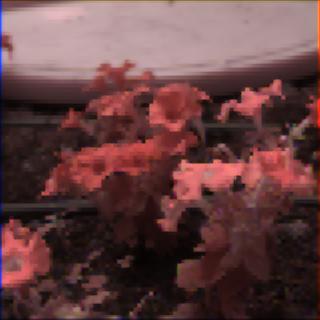} \hspace{\g} &
						\includegraphics[trim={105 185 105 75 },clip,height=\h \textwidth, width=\w \textwidth]{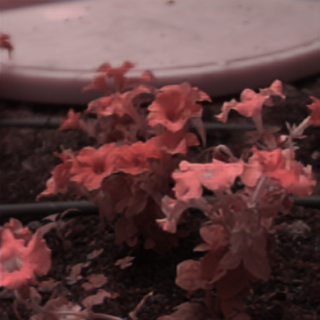} \hspace{\g} &
						\includegraphics[trim={105 185 105 75 },clip,height=\h \textwidth, width=\w \textwidth]{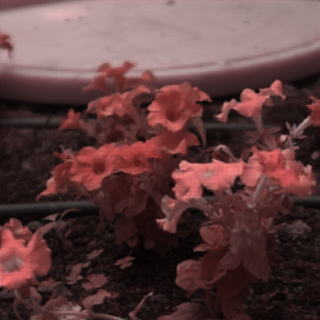}
						\hspace{\g} &		
						\includegraphics[trim={105 185 105 75 },clip,height=\h \textwidth, width=\w \textwidth]{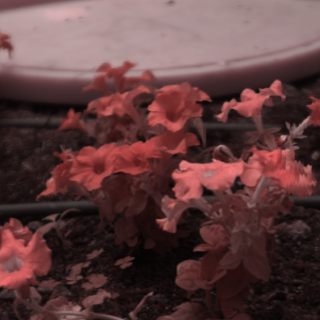} 
						\\ 
						GRMR  \hspace{\g} &	InNet  \hspace{\g} & MCAN
						&
						\textbf{FDM-Net} (ours)
						\\
					\end{tabular}
				\end{adjustbox}
			\end{tabular}	
		\end{tabular}
	}
	\caption{Visual comparison of \textbf{HSI demosaicing} methods (False color, R: 2, G: 11, B:16).} %
	\label{fig_ARAD_903}
\end{figure*}

\newpage

{\small
\bibliographystyle{ieee_fullname}
\bibliography{reference}
}

\end{document}